\documentclass[aps,prb,
reprint,
showpacs,
floatfix,
]{revtex4-1}
\usepackage{amsmath}    
\usepackage{cases} 
\usepackage{graphicx}   
\usepackage{subfigure}
\newcommand{\jun}{junction }
\newcommand{\juns}{junctions }
\newcommand{\Jos}{Josephson }

\begin{document}

\title {Self-field effects in window-type Josephson tunnel junctions}
\author{Roberto Monaco}
\email
{r.monaco@cib.na.cnr.it}
\affiliation{Istituto di Cibernetica del CNR, Comprensorio Olivetti, 80078 Pozzuoli, Italy \\and Facolt$\grave{\rm a}$ di Scienze, Universit$\grave{\rm a}$ di Salerno, 84084 Fisciano, Italy}
\author{Valery P.\ Koshelets}
\affiliation{Kotel'nikov Institute of Radio Engineering and Electronics,
Russian Academy of Science, Mokhovaya 11, Bldg 7, 125009 Moscow, Russia.}
\author{Anna Mukhortova}
\affiliation{Kotel'nikov Institute of Radio Engineering and Electronics,
Russian Academy of Science, Mokhovaya 11, Bldg 7, 125009 Moscow, Russia and \\ Moscow Institute of Physics and Technology, Institutskii per., 9, 
141700 Dolgoprudny, Moscow Region, Russia.}
\author{Jesper Mygind}
\affiliation{DTU Physics, B309, Technical University of Denmark, DK-2800 Lyngby, Denmark}
\date{\today}
\begin{abstract}
The properties of Josephson devices are strongly affected by geometrical effects such as those associated with the magnetic field induced by the bias current. The generally adopted analysis of Owen and Scalapino [{\it Phys. Rev.}{\bf 164}, 538 (1967)] for the critical current, $I_c$, of an in-line Josephson tunnel junction in presence of an in-plane external magnetic field, $H_e$, is revisited and extended to junctions whose electrodes can be thin and of different materials. We demonstrate that the asymmetry of the magnetic diffraction pattern, $I_c(H_e)$, is ascribed to the different electrode inductances for which we provide empirical expressions. We also generalize the modeling to the window-type junctions used nowadays and discuss how to take advantage of the asymmetric behavior in the realization of some superconducting devices. Further we report a systematic investigation of the diffraction patterns of in-line window-type junctions having a number of diverse geometrical configurations and made of dissimilar materials. The experimental results are found in agreement with the predictions and clearly demonstrate that the pattern asymmetry increases with the difference in the electrode inductances.
\end{abstract}

\maketitle
\tableofcontents

\section{Introduction}

\begin{figure}[tb]
\centering
\subfigure[ ]{\includegraphics[width=7.0cm,height=5.1cm]{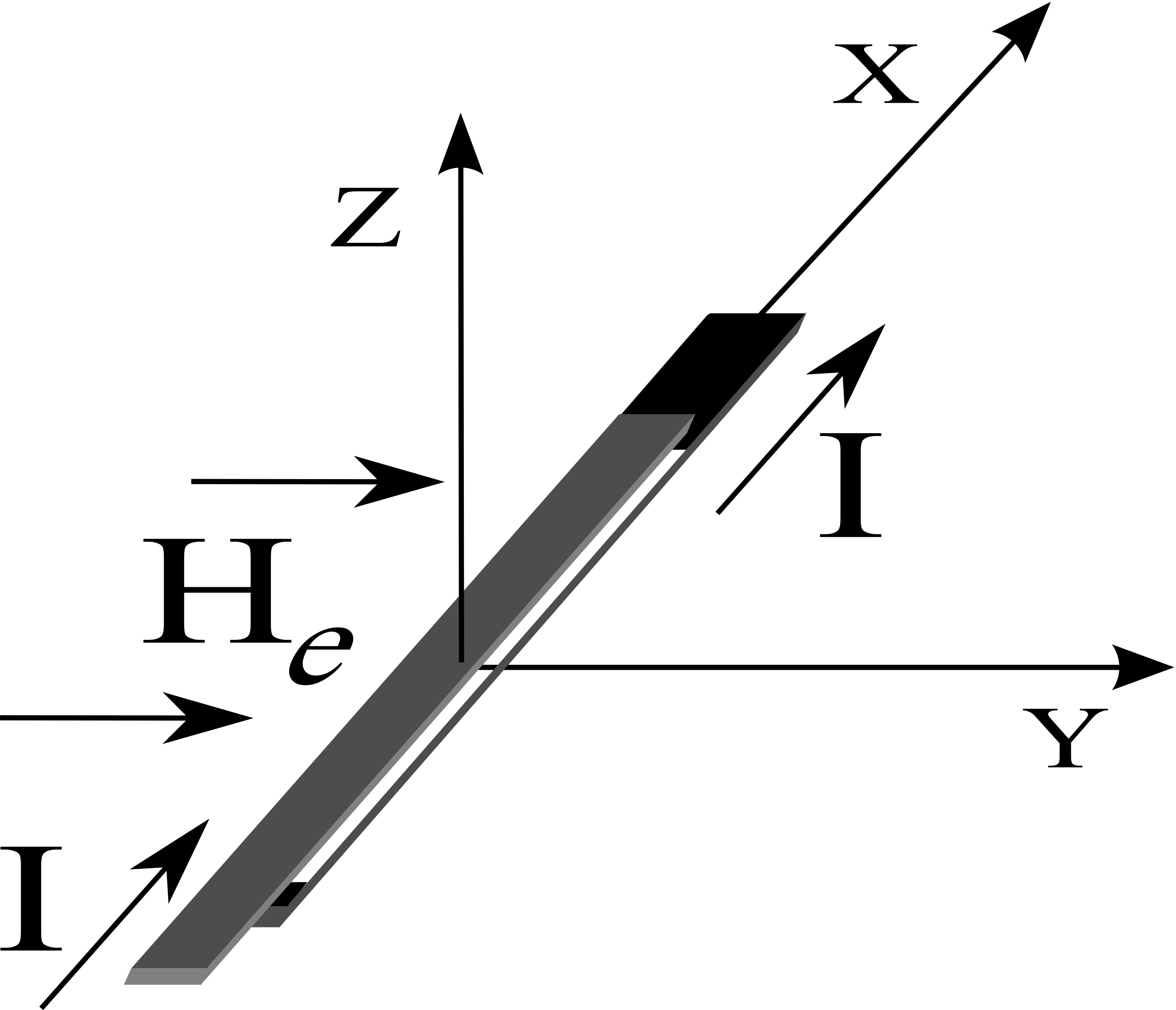}}
\subfigure[ ]{\includegraphics[width=7.0cm,height=3.0cm]{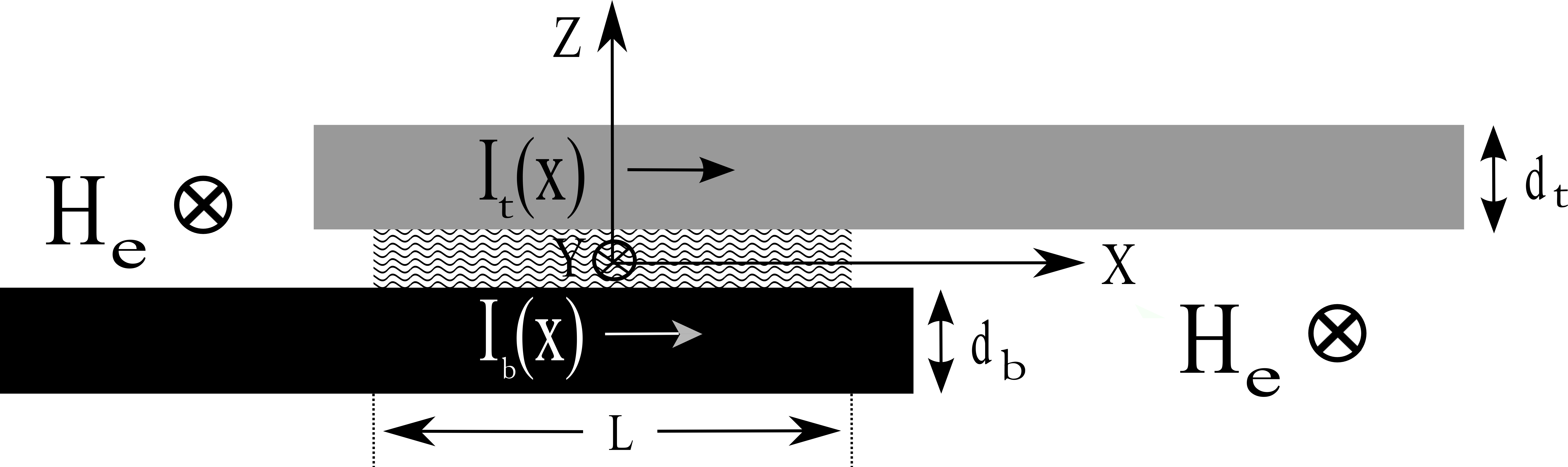}}
\caption{(a) Schematic of an in-line junction with symmetric current bias, $I$, in the presence of an external magnetic field, $H_e$, applied along the $Y$-axis.; (b) Simplified vertical cross section (drawn not to scale) of a in-line Josephson tunnel junction. The junction top electrode is in gray, while the base electrode is in black; the tunneling insulating layer in between is hatched. Also shown is the coordinate system used in this work.}
\label{view}
\end{figure}

In the first years after the discovery of the Josephson effect it was realized that, as the size or the critical current of a planar Josephson tunnel junction increases, the influence of the magnetic field induced by the Josephson current itself becomes more and more important\cite{ferrel,goldman,yamashita,schwidtalA}. Significant advances in the modeling of the so-called \textit{self-field effects} were made by Owen and Scalapino\cite{OS} (OS) who considered the geometrical configuration most suitable to analyze the self-field, namely, the one-dimensional in-line geometry shown in Fig.~\ref{view}(a) where the externally applied bias current, $I$, flows in the direction of the long dimension, $\rm{L}$. The figure also shows the coordinate system used in this work. Long Josephson tunnel junctions (LJTJs), i.e., junctions whose dimension $\rm{L}$ perpendicular to an externally applied magnetic field, $H_e$, is large compared to the Josephson penetration depth, $\lambda_j$, behave like an extreme type-II superconductor; they exhibit a Meissner effect in weak magnetic fields, and vortex penetration starts at a critical field $H_c$. This behavior can be probed into by studying the magnetic field dependence of the maximum tunneling supercurrent, $I_c$, because the Meissner region and the vortex structure are reflected in the $I_c$ versus $H_e$ curve in a very characteristic way. The strength of the current-induced field is measured by the ratio of $\rm{L}/\lambda_j$. The OS analysis was restricted to the ideal case in which the junction electrodes had the same width and were made of the same bulky material. Despite these severe restrictions the OS modeling has been used without further consideration the design of any device based on LJTJs. One of the purpose of the present work is to generalized the OS theory taking into account geometrical or electrical differences in the electrodes. We will extend our interest also to junctions fabricated with the so-called \textit{whole wafer} processes, which make use of a tri-layer structure from which individual junctions are later defined. Due to the large diffusion of these reliable processes, window junctions have become widespread while, in contrast, \textit{naked} (or bare) junctions are a rarity. While the interaction of the Josephson tunnel junction with its embedding circuitry has received an exhaustive and adequate attention\cite{JAP95,franz,caputo,maggi}, the changes due to the self-field effects still remain an unexplored topic, albeit the exact knowledge of the current induced field is highly desirable for the realization of superconductor devices based on LJTJs such as, for example, oscillators\cite{FFO}, magnetic sensors\cite{SUST12} and rectifiers\cite{ratchet}. Besides modeling, a thorough experimental investigation of the self-field effects has been carried out for window-type $Nb$-based LJTJs having in-line configuration and electrodes of different widths, thicknesses and materials. The results are well aligned with the expectations and demonstrate that window-type LJTJs can be designed with a greater flexibility in the control of the self-field effects. The issue of the asymmetry in the threshold curves of Josephson devices was introduced long ago and its importance was recognized in determining the performances of amplifiers\cite{clarke} and magnetometers\cite{tesche}. Currently, the search for new and better ways to achieve and control the asymmetry is still on-going\cite{rudolph,russo}.  

\section{Modeling}

\subsection{Background review}
In its simplest form, the area of a planar Josephson tunnel junction is defined by the overlap of two superconducting films with rectangular cross section and weakly coupled through a thin tunneling barrier. Fig.~\ref{view}(b) depicts the vertical cross section of an in-line Josephson tunnel junction. An in-plane external magnetic field, $H_e$, is applied in the $Y$-direction, i.e., perpendicular to the junction length. The very thin insulating layer between the superconducting films has length $\rm{L}$ and width $W$ (not shown). $I_b(X)$ and $I_t(X)$ denote the local supercurrents flowing, respectively, in the bottom and top junction electrodes within a distance of the order of the penetration depth from the film surfaces and parallel to the insulating layer. $X \in [-{\rm{L}}/2, {\rm{L}}/2]$ is the laboratory spatial coordinate. Throughout the paper the subscripts $b$ and $t$ refer to the base and top electrode, respectively. Further the currents are positive when they flow from the left to the right. In the vast majority of practical cases the junction electrodes are comparable or thinner than their penetration depths and the currents can be well assumed to be uniformly distributed over the film cross section. $I_b$ and $I_t$ also take into account the screening currents, $I_{sc}$, that circulate to expel the magnetic field from the interior of the junction. 

\noindent For in-line LJTJs, it is important to distinguish between the \textrm{symmetric} and \textrm{asymmetric} biasing: in the former, the bias current, $I$, enters at one extremity and exits at the other\cite{OS,stuehm,radparvar81}: $I_t({\rm{L}}/2)=I_b({-\rm{L}}/2)=I$ and $I_t({-\rm{L}}/2)=I_b ({\rm{L}}/2)=0$, while in the latter, the bias current enters and exits from the same extremity\cite{ferrel,stuehm,basa,radparvar85}: $I_t(-L/2)=-I_b(-L/2)=I$ and $I_b(L/2)=I_t(L/2)=0$.

\noindent The gauge-invariant phase difference $\phi$ of the order parameters of the superconductors on each side of the tunnel barrier obeys the Josephson equations\cite{joseph}

\begin{subequations}
\begin{eqnarray}
J_Z(X) = & J_c \sin \phi(X), \label{jos} \\
\kappa {\bf \nabla} \phi(X) & = {\bf H}\times {\bf \hat{n}}, \label{gra}
\end{eqnarray}
\end{subequations}

\noindent in which $J_Z$ is the Josephson current density and $J_c$ is its maximum (or critical) value, which is assumed to be uniform over the barrier area. Eq.(\ref{gra})  states that the phase gradient is everywhere proportional to the local magnetic field ${\bf H}$ and parallel to the barrier plane. Here ${\bf \hat{n}}$ is the versor normal to the insulating barrier separating the superconducting electrodes and $\kappa \equiv {\Phi_0}/{2\pi \mu _0 d_m }$ has the dimension of a current ($\Phi_0$ is the magnetic flux quantum and $\mu_0$ is the vacuum permeability). Neglecting the insulating barrier thickness\cite{wei},

\begin{equation}
\label{dm}
d_m= \lambda_{b} \tanh  \frac{d_b }{2 \lambda_{b}} + \lambda_{t} \tanh \frac{d_t }{2 \lambda_{t}},
\end{equation}

\noindent is the junction \textit{magnetic} thickness in and $\lambda_{b,t}$ and $d_{b,t}$ are, respectively, the bulk magnetic penetration depths and thicknesses of the films. Eq.(\ref{dm}) reduces to $d_m=\lambda_{b}+\lambda_{t}$ in the case of thick superconducting films ($d_{b,t} >4 \lambda_{b,t}$). The net current, $I$, crossing the tunnel barrier is given by

\begin{equation}
I \equiv  W_{} \int_{-{\rm{L}}/2}^{{\rm{L}}/2} J_Z(X)dX.
\label{curr}
\end{equation}

\noindent Throughout this paper we will limit our interest to the zero-voltage time-independent state; this can be achieved as far as $I$ is smaller than the junction critical current, $I_c$. We also assume that the junction width, $W$, is smaller than the Josephson penetration depth\cite{ferrel,joseph,scott76}, $\lambda_J =\sqrt{\Phi_0/ 2\pi \mu _{0}d_j J_{c}}$, setting the length unit of the physical processes occurring in the Josephson junction; here $d_j$ is the junction \textit{current} thickness\cite{wei} (see later)

\begin{equation}
\label{dj}
d_j= \lambda_{b} \coth \frac{d_b }{ \lambda_{b}} + \lambda_{t} \coth \frac{d_t }{ \lambda_{t}} \geq d_m.
\end{equation}

\noindent For thick film junctions, $d_j=d_m=\lambda_{b}+\lambda_{t}$. By its definition, $\kappa \equiv (d_j/d_m) J_c \lambda_J^2$.

\noindent It is well known\cite{ferrel,OS} that combining Eqs.(\ref{jos}) and (b) with the static Maxwell's equations, a static sine-Gordon equation is obtained that describe the behavior of a one-dimensional in-line LJTJ

\begin{equation}
\lambda_J^2 \frac{d^2 \phi}{d X^2} = \sin \phi(X).
\label{sG}
\end{equation}

\noindent Equation(\ref{sG}) was first introduced by Ferrel and Prange\cite{ferrel} in 1963 in the analysis of asymmetrically biased in-line LJTJs; few years later, OS\cite{OS} reported an extensive study of its analytical solutions in terms of elliptic functions for symmetrically biased in-line junctions (provided that ${\rm{L}}\geq\pi \lambda_J /2$). As reported by several authors\cite{stuehm,basa,vaglio}, the largest supercurrent carried by a very long in-line LJTJ is $4I_0$, where $I_0 \equiv J_c W \lambda_J$ is a characteristic junction current (generally from a fraction of a milliampere to a few milliamperes) and depends on the junction normalized length, $\ell\equiv {\rm{L}}/\lambda_J$, as $I_0(\ell)= I_0 \tanh \ell/2$.  

\noindent As already stated, for small fields a LJTJ behaves as a perfect diamagnet by establishing circulating screening currents which maintain the interior field at zero. This Meissner regime is reflected by a linear decrease of $I_c$ with $H_e$. The threshold curves, $I_c(H_e),$ have been the subject of many analytical, numerical and experimental works\cite{barone,vanDuzer,PRB96}. It is well known that the critical magnetic field is\cite{dettmann,franz} $H_c= \Phi_0/\pi \mu_0 \lambda_J d_m = 2J_c \lambda_J d_j/d_m$. The gradual crossover of $H_c$ from short ($\ell<< 2\pi$) to long ($\ell>2\pi$) junctions has been numerically computed\cite{PRB12} and it was found to be well described by the empirical relationship: $H_c(\ell)= H_c \coth \ell/\pi$ in very good agreement with the experimental findings reported in Ref.\cite{dettmann}. Note that in the small junction limit, $ \ell \to 0$, we recover $H_c=\Phi_0/ \mu_0 L d_m$ and the Fraunhofer-like magnetic diffraction pattern (MDP).

\subsection{Boundary conditions}

In this section we will derive the boundary conditions for an in-line junction. Let us define the sheet inductances:

\begin{subequations}
\begin{eqnarray}
T_{b,t} & \equiv & \frac{\mu_0 \lambda_{b,t}}{2} \tanh \frac{d_{b,t}}{2\lambda_{b,t}}; \label{T} \\
C_{b,t} & \equiv & \frac{\mu_0 \lambda_{b,t}}{2} \coth \frac{d_{b,t}}{2\lambda_{b,t}},
\label{C}
\end{eqnarray}
\end{subequations}

\noindent and observe that for any value of the ratio  $d_a/\lambda_{a}$, it is $C_a \geq \mu_0 \lambda_{a}/2 \geq T_a$ with $a=b,t$; in the thin film limit $T_a=\mu_0 d_a/4$ and $C_a=\mu_0 \lambda_a^2/d_a$ is the film kinetic\cite{meservey} sheet inductance due to the inertial mass of mobile charge carriers (Cooper pairs). Furthermore\cite{footnote}, $C_a+T_a= \mu_0 \lambda_{a} \coth\, d_a/\lambda_{a}$, while $C_a-T_a=\mu_0 \lambda_{a} \text{csch}\, d_a/\lambda_{a}$. Then, according to Eqs.(\ref{dm}) and (\ref{dj}), $\mu_0 d_m \equiv 2(T_b+T_t)$ and $\mu_0 d_j \equiv T_b+T_t+C_b+C_t$. 

\noindent In the case of a Josephson structure made by two films of the same width, $W$, Weihnact\cite{wei} provided a general expression for the phase derivative as the sum of three, in general, independent terms

\begin{equation}
\frac{\Phi_0}{2\pi} \frac{d \phi}{d X}= \mu_0 d_m H_e  + \frac{I_t}{W} (C_t+T_b) -  \frac{I_b}{W} (C_b+T_t).
\label{weih}
\end{equation}

\noindent More precisely, if we introduce the magnetic flux changes, $\Delta \Phi_j$, in the Josephson barrier proportional to the changes of the Josephson phase\cite{mc}, $\Delta \Phi_j=\Phi_0 \Delta \phi /2\pi$, each term in Eq.(\ref{weih}) represents a magnetic flux per unit length along the $X$-direction. Let us assume first that the junction is unbiased, so that the screening current, $I_{sc}$, only circulate in the electrodes to expel any external field, $H_e$, from the interior of the junction; charge conservation requires $I_t(X)=-I_b(X)=I_{sc}(X)$. Then:

\begin{equation}
\frac{\Phi_0}{2\pi } \frac{d \phi}{d X}= \mu_0 d_m H_e  +\mu_0 d_j \frac{I_s}{W}.
\label{screening}
\end{equation}

\noindent By resorting to the definition of inductance as flux per unit current\cite{meyers}, we can identify $\mathcal{L}_{JTL} \equiv\mu_0 d_j/W$ as the inductance per unit length for the screening current\cite{scott76}, where the subsript JTL stands for Josephson Transmission Line. The same expression\cite{meyers,swihart,chang,langley} was obtained for a strip-type superconducting transmission line (STL) consisting of a dielectric layer sandwiched between two superconductors of width $W$; here, the current fed into the top electrode returns back in the base electrode acting as a ground plane, namely, $I_b(X)=-I_t(X)$. In such a way, all fields are identically zero outside the waveguide (the only difference being that, for JTLs, the dielectric thickness can be neglected). The screening current in Eq.(\ref{screening}) is negative for positive external field, and vice versa. Since $\mu_0 H_e$ is the externally applied magnetic flux density, $B_e$, and $\mu_0 I_{sc}/W$ is the magnetic flux density, $B_{sc}$, induced in the barrier by the current $I_{sc}$, Eq.(\ref{screening}) help us to understand why we named $d_m$ and $d_j$ as the junction \textit{magnetic} and \textit{current} thickness, respectively. Likewise $\mu_0 d_m$ and $\mu_0 d_j$ can be seen as, respectively, the magnetic and current sheet inductances of the junction. To find the boundary conditions for Eq.(\ref{sG}) when the in-line junction is biased is not an easy task, because it requires the separate knowledge of the properties of each electrode. All previous analytical approaches\cite{swihart,meyers,chang} dealt with transmission line structures in which the current flowing into the top electrode returns back in the lower electrode acting as a ground plane; in such cases the inductance per unit length of the transmission line is the only required parameter. However, for our purposes we have to find the inductances, $\mathcal{L}_b$ and $\mathcal{L}_t$ per unit length of of the bottom and top electrodes, respectively. Resorting to Eq.(\ref{weih}), we easily identify them as:

\begin{equation}
\mathcal{L}_{b,t} \equiv \frac{C_{b,t}+T_{t,b}}{W},
\label{induct}
\end{equation}

\noindent so that $\mathcal{L}_{b}+\mathcal{L}_{t}=\mathcal{L}_{JTL}$: this is a well known identity in the theory of two-conductor transmission lines\cite{kraus,scott76,vanDuzer}. If both films are thick, $\mathcal{L}_b=\mathcal{L}_t =\mu_0(\lambda_b+\lambda_t)/2W$; in the opposite case, $\mathcal{L}_{b,t}=\mu_0 \lambda^2_{b,t}/W d_{b,t}$, that is when both films are thin their inductance is essentially of kinetic origin\cite{meservey}. Classically, the bottom and top inductances have been merged in their parallel combination\cite{scott76,erne80} so that the role played by each supercurrent separately was lost; however, to correctly describe a biased in-line Josephson junction it is mandatory to keep the distinction, since, each electrode transports its own supercurrent, and, in general, $I_b(X) \neq -I_t(X)$. The first term in Eqs.(\ref{induct}) takes into account the magnetic and kinetic energy in the film carrying the current, while the second term is the magnetic energy stored in the opposite electrode. For a thin-film (symmetric) transmission line the ratio of the internal magnetic to the kinetic energy goes\cite{vanDuzer} as $(2d/\lambda)^2/12$, i.e., for thin films the kinetic energy dominates. At this stage we are still allowed to neglect the magnetic energy stored in the very thin insulating tunnel barrier. It is worth to  stress that Eqs.(\ref{induct}) are valid as far as the energy of the magnetic fields outside the transmission line is negligible.

\noindent The case of a superconducting strip of width $W$ carrying a current $I$ over a a superconducting shield was considered a long ago\cite{newhouse}. By resorting to the theorem of images, it was shown that, if the strip is close to a thick superconducting shield, the field between them is approximately uniform and equals to $H=I/W$ and, outside the edge of the film, it falls to zero within a distance of the order of the strip-to-shield distance. The shield screening current, whose integral is equal to $I$, must therefore be uniformly distributed over that portion of the shield surface covered by the film. On the other side of the current-carrying strip the field is zero. In the absence of a shield, it is easy to show that the field on both sides of a freestanding strip will be equal and opposite and of mean magnitude $H=I/2W$. Hence, by bringing a superconducting plane close to a current-carrying strip, we double the field between the plane and the strip and reduce it everywhere else. This effect can be used to reduce the effective inductance of superconducting elements by depositing them on top of another insulated superconductor. These considerations allow the use of Eqs.(\ref{induct}) when a current locally flows either in the bottom or in the top junction electrode, as it occurs in a symmetrically biased in-line junction. However, if in a given place $I_b(X)$ and $I_t(X)$ are comparable and flow in the same direction, then a considerable magnetic field is induced in the outer space (unless the films are thin).

\subsection{Symmetric biasing}

To obtain the local magnetic field, $H_Y(X)$, we have to divide Eq.(\ref{weih}) by $\mu_0 d_m$; in particular, at the boundaries of a symmetrically biased LJTJ we have

$$\kappa{\left. \frac{d \phi}{d X} \right|_{X=-\frac{{\rm{L}}}{2}}}\!\!\!\!\equiv\!H_Y\! \left(\!\!-\frac{{\rm{L}}}{2}\right)\!=\! H_{e} -\frac{\mathcal{L}_b I}{\mu_0 d_m}\!=\! H_{e} -\frac{d_j}{d_m} \frac{\mathcal{L}_b}{\mathcal{L}_{JTL}}\frac{I}{W},$$

\begin{equation}
\kappa{\left. \frac{d \phi}{d X} \right|_{X=\frac{{\rm{L}}}{2}}}\!\!\!\!\equiv\!H_Y\!\!\left( \frac{{\rm{L}}}{2}\right)\!=\! H_{e} + \frac{\mathcal{L}_t I}{\mu_0 d_m}\!=\!H_{e} +
\frac{d_j}{d_m} \frac{\mathcal{L}_t}{\mathcal{L}_{JTL}}\frac{I}{W}.
\label{bc}
\end{equation}

\noindent For $\mathcal{L}_b=\mathcal{L}_t= \mathcal{L}_{JTL}/2$ (and $d_j=d_m$), we recover the symmetric OS boundary conditions\cite{OS} that were generally adopted thereafter for untrue symmetry reasons. In the early eighties\cite{goldman,radparvar81}, the reported asymmetric behavior of samples that were believed to be symmetric led many experimentalists to abandon the in-line geometry in favor of the overlap one. Indeed, the last term in Eqs.(\ref{bc}) is the magnetic field induced in the tunnel barrier by the current $I$ flowing in the lower or in the upper electrodes.

\noindent Ampere's law applied along the barrier perimeter in the $X$-$Y$ plane requires that the magnetic fields at the two ends of the junctions differ by the amount of the enclosed current: $I=W_{} [H_Y ({\rm{L}}/2) -H_Y(-{\rm{L}}/2)]$. It is easy to show that Eqs.(\ref{jos}), (\ref{curr}), (\ref{sG}) and  (\ref{bc}) fulfill Ampere's law. From the boundary conditions (\ref{bc}), it follows that

$$H_Y\left(\!-\frac{{\rm{L}}}{2}\right) + H_Y\left( \frac{{\rm{L}}}{2} \right) = 2 H_{e} + \frac{d_j}{d_m} \frac{I}{W} \frac{\mathcal{L}_t -\mathcal{L}_b }{\mathcal{L}_{JTL}}.$$ The last term, vanishing when $\mathcal{L}_b=\mathcal{L}_t$, has been omitted in all previous analysis of LJTJs. The difference, $\mathcal{L}_t -\mathcal{L}_b$, in the self-inductances of the upper and lower junction electrodes is responsible of the distortion of the experimental curve\cite{goldman}; it equals $\mu_0 [\lambda_{t}\text{csch}\, (d_t/\lambda_{t})-\lambda_{b}\text{csch}\, (d_b/\lambda _{Lb})]$ and reduces to $\mu_0 (\lambda_{t}-\lambda_{b})$ in case of thick electrodes. 
We identify the dimensionless parameter $\alpha \equiv(\mathcal{L}_t -\mathcal{L}_b )/\mathcal{L}_{JTL}$ as a direct measure of the asymmetry of the system, with $-1 \leq \alpha \leq 1$. The asymmetry can be ascribed to differences in the electrode thicknesses and/or materials, that is, it can have geometrical and/or electric origins. 

\noindent We like to point out that Eqs.(\ref{bc}) are very general and can be used to correctly describe the so-called {\it self-field effects} occurring in any {\it substantially} in-line LJTJ in which the bias current or a part of the bias current enters or leaves the junction from one or even both its ends. Unfortunately, their implementation requires the separate knowledge of the bottom and top electrode inductances per unit length (rather than just their sum). 

\subsection{Asymmetric biasing}

In the case of asymmetric bias, the boundary conditions are
$$ H_Y\left(\! -\frac{{\rm{L}}}{2}\right) =\! H_e - \frac{d_j}{d_m} \frac{\mathcal{L}_b+\mathcal{L}_t }{\mathcal{L}_{JTL}}\frac{ I}{W},$$

\begin{equation}
H_Y\left(\frac{{\rm{L}}}{2}\right) = H_e,
\label{bca}
\end{equation}

\noindent and, recalling that $\mathcal{L}_b+\mathcal{L}_t= \mathcal{L}_{JTL}$, we end up with the boundary conditions first found by Ferrel and Prange\cite{ferrel} for thick electrode LJTJs. We observe that the difference in the inductances does not play any role in this case. Many pioneeristic experiments\cite{basa,matisoo} were also carried out with in-line junctions built over a large and thick superconducting ground plane that makes the current $I_b$ to flow in the lower skin layer of the bottom electrode provided it is a thick film and is shielded by the upper part. When this is the case, $\mathcal{L}_b$ must be set to zero in Eqs.(\ref{bc}) or (\ref{bca}), so that the biasing configuration becomes irrelevant. For the above reasons, in the rest of this paper we will only consider the more interesting case of the symmetrically biased in-line junction with no ground plane.

\subsection{Normalized units}

\noindent In normalized units of $x\equiv X/\lambda_J$, the differential equation (\ref{sG}) becomes:

\begin{equation}
\phi_{xx} = \sin \phi(x), 
\label{ODE}
\end{equation}

\noindent with $x \in [-\ell/2,\ell/2]$. Normalizing the currents to $I_0 \equiv J_c W_{} \lambda_J$ and the magnetic fields to $H_c/2 \equiv J_c \lambda_J d_j/d_m$, the boundary conditions (\ref{bc}) for a symmetrically biased LJTJ read

\begin{equation}
\phi_{x}\!\left(\!\!-\frac{\!\ell}{2}\!\right) \!\equiv h_l\!= h_e - \frac{\mathcal{L}_b\, i}{\mathcal{L}_{JTL}};  \,\,\,   \phi_{x}\!\left(\!\frac{\ell}{2}\!\right)\!\equiv  h_r \!=  h_e + \frac{\mathcal{L}_t\,i}{\mathcal{L}_{JTL}}.  
\label{bcn}
\end{equation}

\noindent With these notations, the normalized critical magnetic field $h_c$ of a short \Jos \jun is $2\pi/\ell$. 

\subsubsection{Magnetic diffraction pattern}

Setting $h_l$ and $h_r$ at their extreme values $\pm2$ in Eqs.(\ref{bcn}), we obtain the  normalized MDP, $i_c(h_e)$, in the Meissner regime

\begin{equation}
    i_c(h_e) = 
    \begin{cases}
       \frac{2+h_e}{{\mathcal{L}_b}/{\mathcal{L}_{JTL}}} & \mbox{for $-2 \leq h_e \leq h_{max}$} \\
       \frac{2-h_e}{{\mathcal{L}_t}/{\mathcal{L}_{JTL}}} & \mbox{for $h_{max} \leq h_e \leq 2,$}
    \end{cases}
\label{mdp}    
\end{equation}

\noindent with $h_{max}= 2\alpha$ being the field value which yields the maximum critical current $i_c(h_{max})=4$. The second-last equality turns out to be very useful in the experiments to determine the asymmetry parameter from the analysis of the junction MDP:

\begin{equation}
\alpha = \frac{h_{max}}{2}=\frac{H_{max}}{H_c}.
\label{alpha}
\end{equation}

\begin{figure}[tb]
\centering
\includegraphics[width=7.0cm,height=6cm]{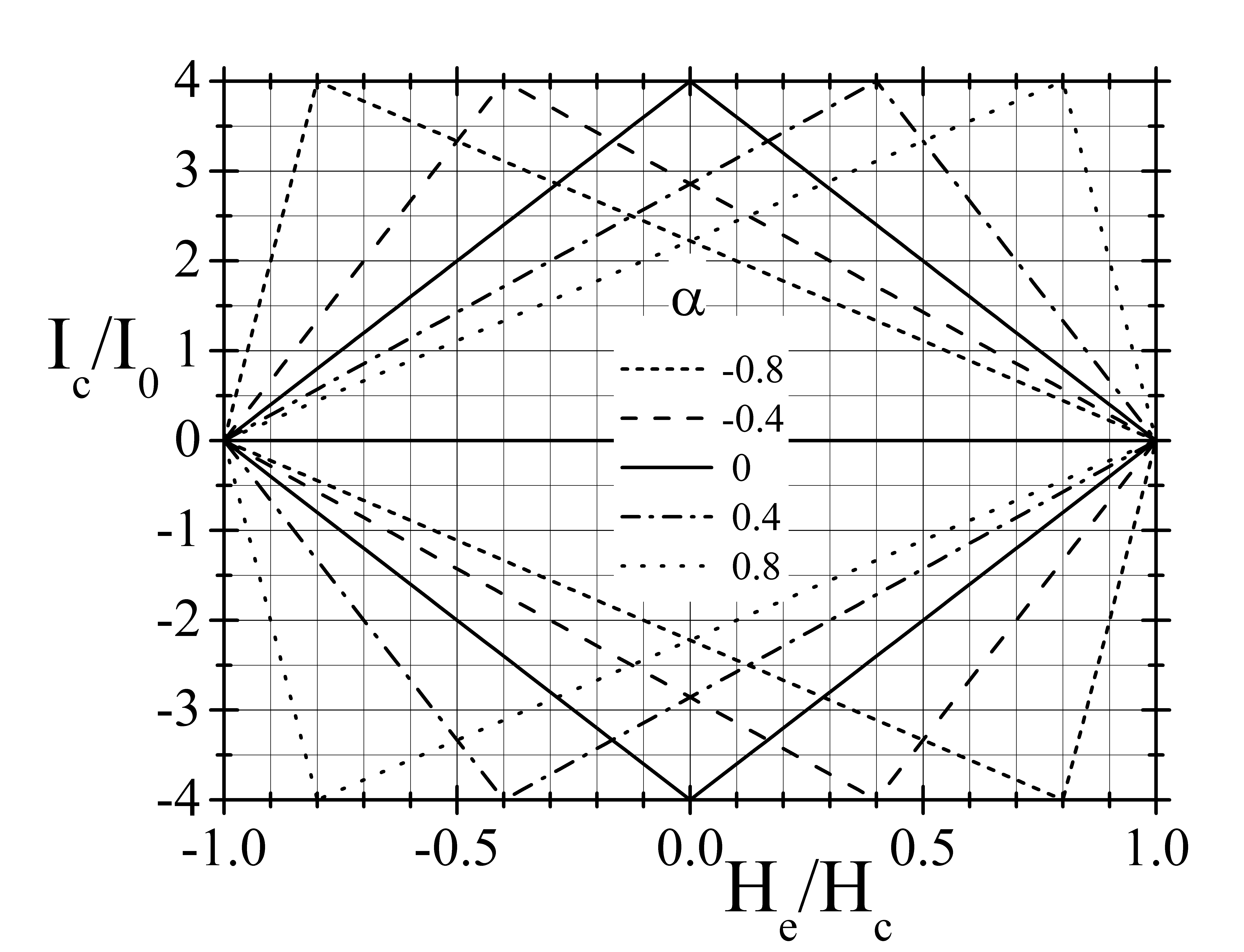}
\caption{Theoretical magnetic diffraction patterns $I_c(H_{e})$ of a very long in-line symmetrically current-biased Josephson junction in the Meissner regime for different values of $\Lambda_t \equiv \mathcal{L}_t/\mathcal{L}_{JTL}$. The critical current, $I_c$, is normalized to $I_0=J_c W \lambda_J$ and the externally applied magnetic field, $H_e$, to the critical magnetic field $H_c$.}
\label{mdpp}
\end{figure}

\noindent Fig.~\ref{mdpp} shows the theoretical $I_c(H_{e})$ for a very long in-line Josephson junction in the Meissner regime and for different values of $\alpha$; the junction critical current, $I_c$, is normalized to $I_0$ and the critical magnetic field, $H_c$, is the theoretical field value that fully suppresses the critical current. We remark that the patterns are antisymmetric, $I_c(-H_e)=-I_c(H_e)$ and piecewise linear. They have different absolute slopes $|dI_c/dH_e|$ on the left and right branches, respectively, equal to $\mu_0 d_m /{\mathcal{L}_b}$ and $\mu_0 d_m /{\mathcal{L}_t}$. 

\subsubsection{Current diffraction pattern}

Let us consider now the case when the junction critical current is modulated by the transverse field induced by a stationary current entirely flowing in either the top or bottom electrode. This so-called \textit{control current} technique has been used to produce local magnetic fields for digital applications of Josephson circuits since 1969\cite{matisoo}. If the control current, $\hat{I}$, is injected into, say, the top electrode, then the magnetic field at the inner surfaces is $\hat{H}_t= \mathcal{L}_t \hat{I} /\mu_0 d_m$, as in Eq.(\ref{bc}). The value of the control current for which the junction critical current $I_c$ vanishes is $\hat{I}_c=\Phi_0/\pi \lambda_J \mathcal{L}_t$ and will be named the critical control current. Likewise, $\hat{I}_{max}$ will be that value of $\hat{I}$ which maximizes the critical current, $I_c$. In normalized units, $\hat{h}_t \equiv 2 \hat{H}_t/H_c = \mathcal{L}_t \hat{\iota}/\mathcal{L}_{JTL}$, where $\hat{\iota} \equiv \hat{I}/I_{0}$. Replacing $h_e$ with $\hat{h}_t$, Eqs.(\ref{mdp}) become

\begin{equation}
\label{mdpcl}
i_c(\hat{\iota})=
	\begin{cases}
			\frac{ \mathcal{L}_t }{ \mathcal{L}_b} (\hat{\iota}_{c} + \hat{\iota}) & \mbox{for $-\hat{\iota}_{c} \leq \hat{\iota} \leq \hat{\iota}_{max}$}\\
			\hat{\iota}_{c} - \hat{\iota} & \mbox{ for $\hat{\iota}_{max} \leq \hat{\iota} \leq \hat{\iota}_{c},$}\\
	\end{cases}
\end{equation}

\noindent where $\hat{\iota}_{c} \equiv \hat{I}_c/I_0 = 2 \mathcal{L}_{JTL}/ \mathcal{L}_t$  and $\hat{\iota}_{max} \equiv \hat{I}_{max}/I_0 = \hat{\iota}_{c}(\mathcal{L}_b -\mathcal{L}_t)/\mathcal{L}_{JTL}$. It is worth mentioning that

\begin{equation}
\frac{\hat{I}_{max}}{\hat{I}_c}=\frac{H_{max}}{H_c}
\label{Icc}
\end{equation}

\noindent indicating that the degree of asymmetry is the same for the magnetic and the current diffraction pattern (CDP). In other words, as far as the modulation of the critical current concerns, a control current is equivalent to an externally applied field. It is then possible to identify the CDP slope with a current gain\cite{clarke71a,JAP11} $g \equiv dI_c/d\hat{I}= di_c/d \hat{\iota}$. From Eq.(\ref{Icc}), it follows

\begin{equation}
\label{gain}
g=
\begin{cases} 
			 \frac{\mathcal{L}_t}{\mathcal{L}_b} & \mbox{for $-\hat{\iota}_{c} \leq \hat{\iota} \leq \hat{\iota}_{max}$}\\
			-1 & \mbox{for $\hat{\iota}_{max} \leq \hat{\iota} \leq \hat{\iota}_{c}.$}
\end{cases}
\end{equation}

\noindent With $\mathcal{L}_t > \mathcal{L}_b$, then $\hat{\iota}_{max}$ is negative and so, in zero external field, we have a unitary current gain. However, large current gains can be achieved with samples having a large $\mathcal{L}_t/ \mathcal{L}_b$ ratio by control current biasing the upper electrode to have $\hat{\iota} < \hat{\iota}_{max}$.   

\section{Window-type junctions}

Weihnact's Eq.(\ref{weih}) was derived considering that the thickness of the tunnel barrier is much smaller that typical penetration depths so that the magnetic field in between the electrodes is in the $Y$-direction, is uniform along $Z$, and only depends on the $X$ coordinate. Accordingly, there are no fringing effects and, at the same time, the magnetic energy stored in the dielectric layer can be neglected. In addition, it was assumed that the base and top electrode match the width of the barrier, $W_b=W_t=W$. However, these conditions are not fulfilled in nowadays window-type planar tunnel junctions whose electrodes have quite different widths; typically, $W_b>W_t>W_{}$, that is, two strips, of total width $W_i=W_t-W$, exist along the junction sides where the top electrode overhangs the base electrode, but no tunneling is possible due to the thick insulating layer. In case $W_t>W_b$, then $W_i=W_b-W$; however, throughout this paper we will assume that the base electrode is wider or at most matches the top electrode. The so-called \textit{idle} region is formed after the patterning of the wiring film which provides the electrical connection to the junction top electrode. In this region surrounding the tunnel area the insulation between the bottom and top electrode is provided by an oxide layer typically made of a native anodic oxide of the base electrode and/or a deposited $SiO_x$ layer. The total thickness of this passive layer, $d_{ox}$, is comparable or even larger than the electrode penetration depths, $\lambda_{b,t}$, and might also be comparable with the junction width, $W$. Then the screening currents in the upper and lower electrodes distribute over an effective width larger than the junction width. This leads to a new smaller junction inductance per unit length $\mathcal{L'}_{JTL}$ corresponding to the so-called\cite{JAP95,franz,caputo} \textit{inflation} of the Josephson penetration depth occurring in window-type \Jos tunnel junctions. In Refs.\cite{JAP95,franz}, each electrode was modeled as a parallel combination of two stripes having quite different oxide thicknesses resulting in a rather involved expression of the effective current thickness. From a phenomenological point of view, a unidimensional junction with a lateral idle region behaves as a bare junction having the same width, $W$, and length, $\rm{L}$, but with an effective current thickness, $d'_j$, given by a proper combination of the current thicknesses of the naked junction, $d_j$ and of the idle region, $d_i$, namely: $d'_j= d_j/[1+(d_j/d_i)(W_i/W_j)]<d_j$, so that $\mathcal{L'}_{JTL} \equiv \mu_0 d'_j/W$ and $\lambda'_J \equiv \sqrt{\Phi_0/ 2\pi \mu _{0}d'_j J_{c}}>\lambda_J$. The prime symbol (') labels the parameters relative to the window junction. Taking into account the thickness, $d_w$, of the wiring layer, it is: 
 
$$d_j= \lambda_{b} \coth \frac{d_b }{ \lambda_{b}} 
+ \lambda_{t} \coth \frac{(d_t+d_w)}{ \lambda_{t}},$$ 

$$d_i= \lambda_{b} \coth \frac{d_b }{ \lambda_{b}} 
+ \lambda_{t} \coth \frac{d_w }{ \lambda_{t}} + d_{ox}. $$ 

Outside the junction area the idle region takes the form of a microstrip-line made by two electrodes of finite width and thickness. Chang\cite{chang} considered the case of a superconducting strip transmission line, i.e., a structure consisting of a finite-width superconducting film of thickness $h$ over an infinite (and thick) superconducting ground plane. As far as the strip linewidth, $w$, exceeds about the insulation thickness, $t$, the inductance per unit length, $\mathcal L_{STL}$, was analytically derived as 

\begin{equation}
\mathcal{L}_{STL}=\frac{\mu_0 d_j}{w}K^{-1}\left(\frac{w}{h},\frac{t}{h}\right), 
\label{LS}
\nonumber
\end{equation}

\noindent with the fringing-field functional $K$, first introduced in Ref.\cite{newhouse},  being always larger than unity; the fringing fields have the effect to increase the system energy, i.e., to reduce the inductances per unit length. $K$ decreases with the the ratio $w/h$ and increases with $t/h$. Definitely, Chang's results can be used when $W_b >> W_t$, but, unfortunately, no analytical expression is available for a STL when both electrodes have finite and comparable widths. 

\noindent The situation complicates if $I_b(X) \neq - I_t(X)$. Therefore, we resort again to Eq.(\ref{weih}) to determine the flux per unit length just outside the tunneling area

\begin{equation}
\frac{\Phi_0}{2\pi} \frac{d \phi}{d X}= \mu_0  d'_m  H_e+  \mathcal{L'}_t I_t - \mathcal{L'}_b I_b.
\label{weih'}
\end{equation}

\noindent In other words, for a window-type LJTJ the left (right) boundary condition is determined by the property of the longitudinal idle region to the left (right) of the left (right) junction end.  The longitudinal idle region also acts on the junction as a capacitive load, which does not play any role, as far as the static properties concern. We propose that, provided that $W_b$ slightly exceeds $W_t$, the inductances per unit length of the bottom and top electrodes at the extremities of a symmetrically biased in-line junction (where the bias current, $I$, at the junction extremities either flows is the bottom or in the top electrode) are

\begin{subequations}
\begin{eqnarray}
\mathcal{L'}_{t} & \approx & \frac{C_{t}+T_{b}+ \mu_0 d_{ox}}{K_t W_{t}},
\label{L't}\\
\mathcal{L'}_{b} & \approx & \frac{C_{b}+T_{t}+ \mu_0 d_{ox}}{K_b W_{b}}.
\label{L'b}
\end{eqnarray}
\end{subequations}

\noindent where $K_{b,t}$ are fringing field factors that remain to be determined: with $W_b\geq W_t$ we expect $K_b\geq K_t\approx 1$. Eqs.(\ref{L't}) and (b) rely on the fact that a current $I_t$ ($I_b$) flowing in the top (bottom) electrode induces a magnetic field $I_t/W_t$ ($I_b/W_b$) between the electrodes. The last terms in each of the previous equations take into account the magnetic energy stored in the thick oxide layer. In Eq.(\ref{weih'}) we also have considered that the magnetic thickness outside the tunneling area is $d'_m=d_m+ d_{ox}$ (but this is a secondary field focusing effect\cite{vanDuzer,lee}). Consequently, for any window-type Josephson tunnel junction the critical field is lower than that of a naked one; for a window-type LJTJ it is

\begin{equation}
H'_c= \Phi_0/\pi \mu_0 d'_m \lambda'_J < H_c.
\label{Hc'}
\end{equation}

\noindent Ultimately, the boundary conditions Eqs.(\ref{bc}) still hold, if we replace $d_m$, $d_j$, $\mathcal{L}_{b,t}$ and $\mathcal{L}_{JTL}$ for the naked junction with their respective counterparts, $d'_m$, $d'_j$, $\mathcal{L'}_{b,t}$ and $\mathcal{L'}_{JTL}$, for the window junction. It should be clear that, this time, the inductance per unit length, $\mathcal{L'}_{STL}$, of the superconducting (non-Josephson) transmission line outside the Josephson area is not given by the sum $\mathcal{L'}_{b}+\mathcal{L'}_{t}$, otherwise the magnetic energy stored in the dielectric layer would be counted twice. Naturally, in general, $\mathcal{L'}_{b}+\mathcal{L'}_{t}$ not even matches $\mathcal{L'}_{JTL}$. For the samples fabricated by means of a tri-layer process, the top electrode to be considered in Eqs.(\ref{L't}) and (b) is the wiring layer which does not necessarily cover the whole barrier area, so that it could as well be $W_t<W$.

Indeed, the in-line approximation fails, if $W_b$ or $W_t$ are much larger than $W$, because part of the bias current enters the tunnel barrier also along the long junction dimension, $\rm{L}$, and a mixed in-line-overlap model\cite{olsen,sarnelli} should be adopted in which the Josephson phase $\phi$ obeys the time-independent perturbed sine-Gordon equation

\begin{equation}
\nonumber
\sin \phi(X) = \lambda_J^2 \frac{d^2 \phi(X)}{d X^2}+ \frac{\mathcal{I}_{ov}(X)}{J_c W}. 
\label{sG'}
\end{equation}

\begin{figure}[tb]
\centering
\subfigure[ ]{\includegraphics[width=7cm]{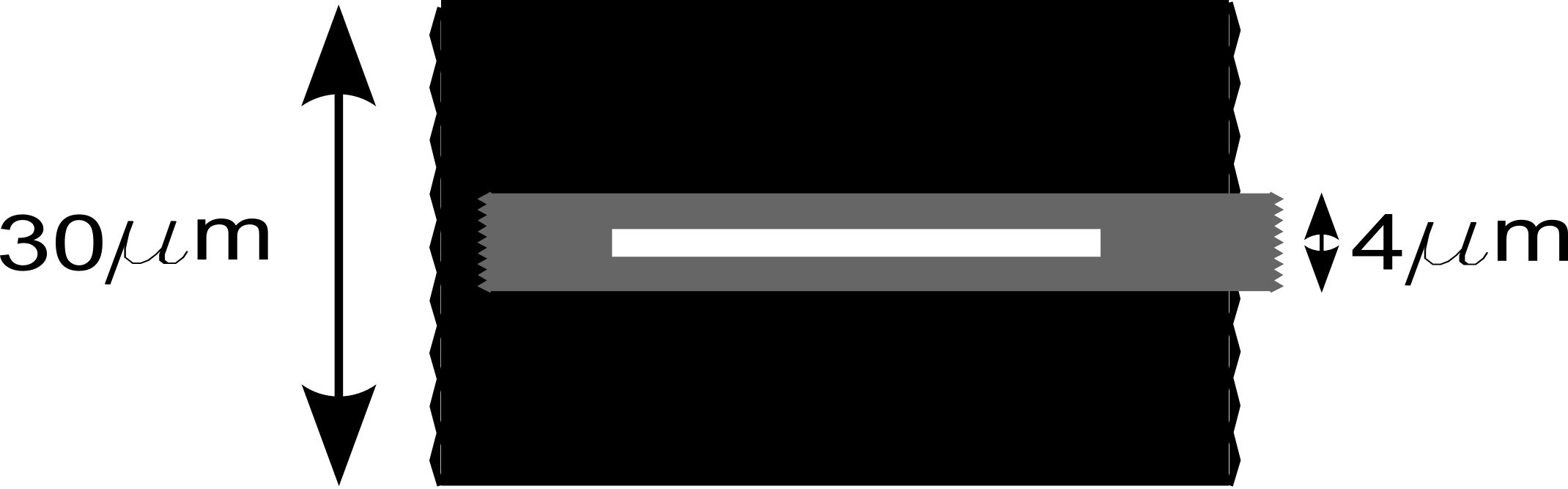}}
\subfigure[ ]{\includegraphics[width=7cm]{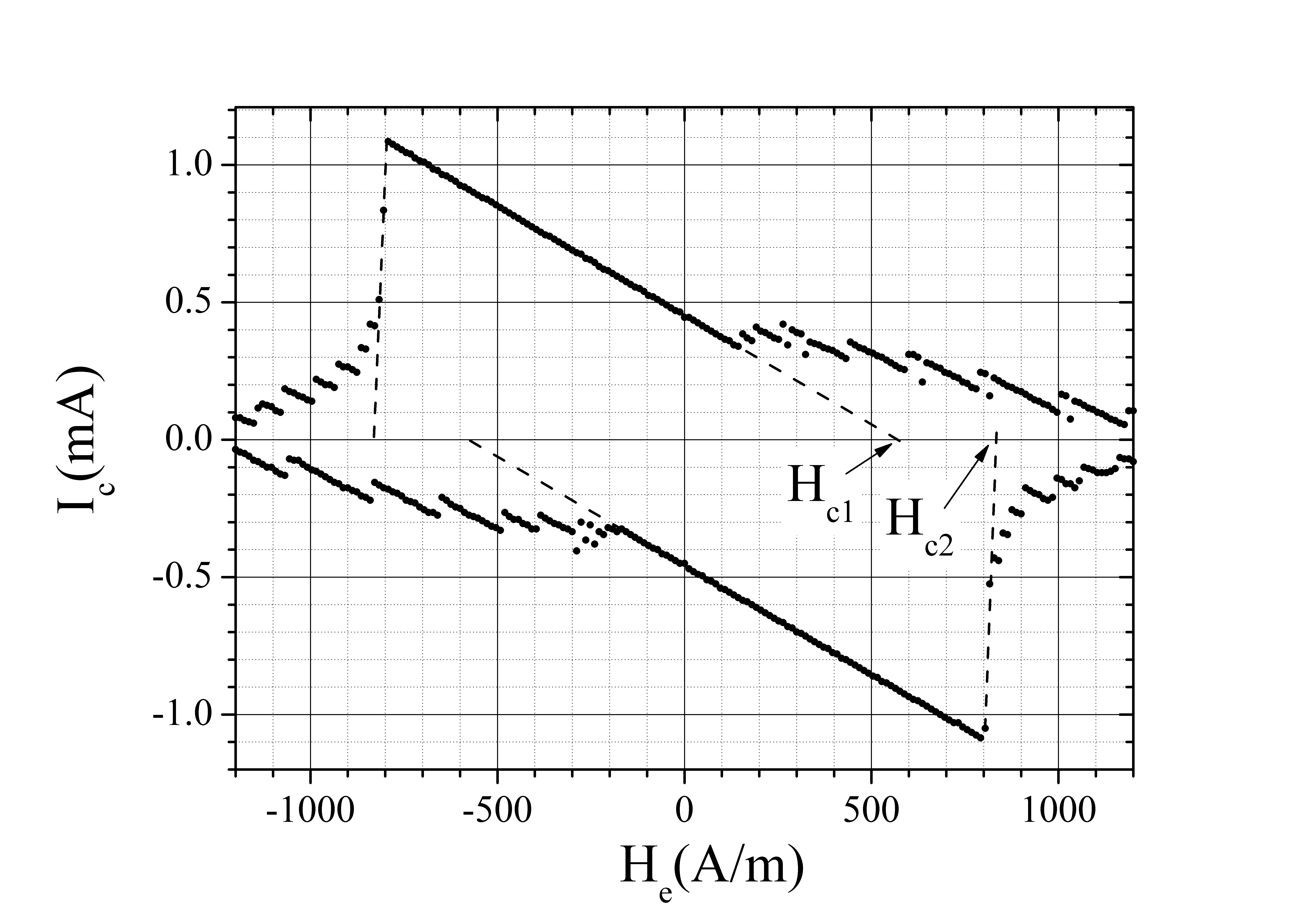}}
\caption{(a) Electrode configuration of a $100\, \mu$m long and $1.2\, \mu$m wide, window-type Josephson tunnel junction having quite different electrode widths: $W_t=4\, \mu$m and $W_b=30\, \mu$m; (b) Its experimental magnetic diffraction patterns, $I_c(H_{e})$, when symmetrically biased: $I_{c,max}=1.10\, mA$, $H_{c1}=\pm 580\, A/m$ and $H_{c2}=\pm 830\, A/m$.}
\label{wide}
\end{figure}

\noindent $\mathcal{I}_{ov}(X)$ is the distribution of the externally applied bias current, $I_{ov}=\int^{L/2}_{-L/2}\mathcal{I}_{ov}(X) dX$, giving the fraction of the bias current $I$ that enters the junction through the long dimension (overlap component). Nevertheless, the junction configuration is still substantially in-line and, as such, the boundary conditions  discussed so far still apply with the only caveat that the in-line component of the bias current is $I-I_{ov}$. The consequences resulting from the asymmetric boundary conditions imposed by a non-uniform external magnetic field at the extremities of both short and long \Jos \juns have been recently investigated \cite{JAP10,APL11}; the field asymmetry is responsible for a degeneracy of the critical field, $H_c$, that was numerically demonstrated and experimentally verified. To better understand this point, we sketch in Fig.~\ref{wide}(a) the geometry of a $100\, \mu$m long and $1.2\, \mu$m wide, symmetrically biased window-type $Nb$-$AlN$-$NbN$ junction with $W_t=4\, \mu$m and $W_b=30\, \mu$m. Fig.~\ref{wide}(b) shows the MDP in which the two critical fields, $H_{c1}$ and $H_{c2}$, are quite evident. Since in real measurement $I_c$ never vanishes before the next lobe "grows up", the critical field values are obtained by extrapolating to zero the linear branches of the principal lobes; this is indicated by the gray dashed lines in this and forthcoming  plots. A similar pattern asymmetry was first reported in Ref.\cite{schwidtalA} and was erroneously ascribed to a low uniformity of the barrier properties. However, this is an extreme case in which the asymmetry is mainly due to the very different widths of the electrodes. The experimental findings that will be presented in the next Section refer to window junctions having $W_b \geq 1.5 W_t$ for which the degeneracy of the critical fields, although observable, is small enough to be ignored.

\subsection{Normalized units}

\noindent The boundary conditions for a window junction are obtained dividing  Eq.(\ref{weih'}) by $\mu_0 d'_m$. In normalized units of $x'\equiv X/\lambda'_J$, with $x' \in [-\ell'/2,\ell'/2]$ where $\ell' \equiv {\rm{L}}/\lambda'_J$ is the normalized length of the window junction; normalizing the currents to $I'_0 \equiv J_c W_{} \lambda'_J$ and the magnetic fields to $H'_c/2=J_c \lambda'_J d'_i/d'_m$, Eqs.(\ref{bcn}) become

\begin{equation}
\phi_{x'}\!\!\left(\!\!-\frac{\ell'}{2}\!\right)\!\!\equiv h'_l= h'_e - \Lambda_b'  i;  \quad   \phi_{x'}\!\!\left(\!\frac{\ell'}{2}\!\right)\!\!\equiv  h'_r =  h'_e + \Lambda'_t i,  
\label{bcn'}
\end{equation}

\noindent where we have introduced the reduced inductances $\Lambda'_{b,t}\equiv \mathcal{L'}_{b,t}/\mathcal{L'}_{JTL}$ and, in general, $\Lambda'_b + \Lambda'_t \neq 1$. Setting $h'_l$ and $h'_r$ at their extreme values $\pm2$ in Eqs.(\ref{bcn'}), we obtain the MDP $i'_c(h'_e)$ in the Meissner regime lobe

\begin{equation}
    i'_c(h'_e) = 
    \begin{cases}
       \frac{2+h'_e}{ \Lambda'_b} & \mbox{for $-2 \leq h'_e \leq h'_{max}$} \\
       \frac{2-h'_e}{\Lambda'_t} & \mbox{for $h'_{max} \leq h'_e \leq 2,$}
    \end{cases}
\label{mdp'}    
\end{equation}

\noindent with $h'_{max}\equiv 2(\mathcal{L'}_b -\mathcal{L'}_t)/(\mathcal{L'}_b +\mathcal{L'}_t)$ being the normalized field value yielding the maximum critical current $i'_c(h'_{max})=4/(\Lambda'_b +\Lambda'_t)=4\mathcal{L'}_{JTL}/(\mathcal{L'}_b +\mathcal{L'}_t)$. Still we can define the asymmetry parameter 

\begin{equation}
\alpha' \equiv \frac{\mathcal{L'}_t -\mathcal{L'}_b}{\mathcal{L'}_{JTL}}= \frac{H'_{max}}{H'_c}
\label{alpha'}
\end{equation}

\noindent and express the ratio of the inductances per unit length as

\begin{equation}
\frac{\mathcal{L'}_t}{\mathcal{L'}_b}= \frac{1- \alpha'}{1 + \alpha'};
\label{ratioex}
\end{equation}

\noindent when $\alpha'$ is positive then $\mathcal{L'}_t < \mathcal{L'}_b$ and vice versa. If the modulating magnetic field is the transverse field induced by a control current, $\hat{I}$, flowing in the top electrode, then $\hat{H'}_t= \mathcal{L'}_t \hat{I} / \mu_0 d'_m = d'_j \Lambda'_t \hat{I}/d'_m W$. In normalized units,  $\hat{h'}_t \equiv 2\hat{H'}_t/H'_c= \Lambda'_t \hat{\iota}'$, where $\hat{\iota}' \equiv \hat{I}/I'_{0}$ is the reduced control current. Replacing $h'_e$ with $\Lambda'_t \hat{\iota}'$ in Eq.(\ref{mdp'}), we get the CDP of a window-type junction

\begin{equation}
\label{mdpcl'}
i'_c(\hat{\iota}')=
	\begin{cases}
				\frac{2+\Lambda'_t \hat{\iota}'}{ \Lambda'_b} & \mbox{for $-\hat{\iota}'_{c} \leq \hat{\iota}' \leq \hat{\iota}'_{max}$}\\
				\frac{2-\Lambda'_t \hat{\iota}'}{\Lambda'_t} & \mbox{ for $\hat{\iota}'_{max} \leq \hat{\iota}' \leq \hat{\iota}'_{c},$}\\
	\end{cases}
\end{equation}

\noindent where $\hat{\iota}'_{c} \equiv 2/\Lambda'_t$  and $\hat{\iota}'_{max} \equiv h'_{max}/\Lambda'_t = 2(\Lambda'_b/\Lambda'_t -1)/(\Lambda'_b+\Lambda'_t)$. The current gain, $g' \equiv  di'_c/d \hat{\iota}'$, for a in-line window-type junction is

\begin{equation}
\label{gain'}
g'=
\begin{cases} 
			\frac{\Lambda'_t}{\Lambda'_b}=\frac{\mathcal{L'}_t}{\mathcal{L'}_b}   & \mbox{for $-\hat{\iota}'_{c} \leq \hat{\iota}' \leq \hat{\iota}'_{max}$}\\
			-1 & \mbox{ for $\hat{\iota}'_{max} \leq \hat{\iota}' \leq \hat{\iota}'_{c}.$}\\
\end{cases}
\end{equation}

\noindent Since most real samples have $W_b>W_t$, in the case of electrodes having the same penetration depth, then $\mathcal{L'}_t>\mathcal{L'}_b$, suggesting that, for applications it is preferable to inject the control current into the top electrode which, typically, with respect to the bottom electrode, has a smaller width and so a larger inductance per unit length. Furthermore, $\mathcal{L'}_t$ can be increased by simply reducing the wiring width at the junction extremities. This can be an advantage
for some applications.

\section{Experiments}

The findings of the previous Section can be experimentally verified by measuring the magnetic and current diffraction patterns of symmetrically biased in-line window-type LJTJs. In particular, Eqs.(\ref{alpha'}) and (\ref{ratioex}) allow the determination of the ratio of the inductances per unit length which can be compared with the expected value from Eqs.(\ref{L't}) and (b):

\begin{equation}
\frac{\mathcal{L'}_{t}}{\mathcal{L'}_{b}}  = \sigma \frac{W_{b}}{ W_{t}} \frac{C_{t}+T_{b}+  \mu_0 d_{ox}}{C_{b}+T_{t}+  \mu_0 d_{ox}}
\label{ratioth}
\end{equation}

\noindent where $\sigma \equiv K_b/K_t$ is a factor which takes into account the film asymmetry \cite{langley} and is equal to $1$ when the electrodes have the same geometrical and electrical parameters. We will test the validity of the Eq.(\ref{ratioth}), for samples having $W_b =W_t$ and $W_b =1.5 W_t$.
 
\subsection{Samples}

In order to have a solid benchmark of data, we have compared the static properties of thin-film $Nb$-$Nb$ and $Nb$-$NbN$ samples having the same geometrical configuration. The $NbN$ films used in this study were deposited by dc magnetron sputtering of a pure $Nb$ target in an argon and nitrogen sputtering gas. Although the samples were not heated intentionally, the substrate temperature slightly rose during deposition; nevertheless surface temperature never exceeds $120\,^o$C allowing for a lift-off process. The superconducting transition temperature (determined as the resistivity midpoint by a four-point method) for samples deposited under optimal conditions is $15.5\,$K. Polycrystalline Niobium Nitride has a dirty-limit penetration depth\cite{oates,villegier}, $\lambda_{NbN}(T=4.2K)=370\,nm$, several times larger than that of epitaxially grown Niobium, $\lambda_{Nb}(T=4.2K)=90\,nm$.

\noindent All samples were long window-type in-line junctions having physical length $\rm{L}=100\, \mu m$ and width $W=1.2\,\mu m$ with a $1.4\,\mu m$ wide idle region on each side, so that $W_i=2.8\,\mu m$. They were fabricated with the same parameters for the deposition and anodization of the base electrodes, as well as for the deposition of the passive layer: $d_b=190 \pm 10\,nm$ and $d_{ox}=220 \pm 10\,nm$. Further, the top and wiring layers had quite similar thicknesses: $d_{t,Nb}=d_{t,NbN}=65 \pm 5\,nm$, $d_{w,Nb}=470 \pm 10\,nm$ and $d_{w,NbN}=390 \pm 10\,nm$. Therefore, the material used for the top (and wiring) electrode was the only notable difference. More importantly the wiring layers were $Nb$-$Nb$ samples, $d_{w,Nb} \approx 6 \lambda_{Nb}$, and thin for the $Nb$-$NbN$ ones, $d_{w,NbN} \approx  \lambda_{NbN}$, in which case the inductance is predominantly kinetic. We also note that the thickness, $d_{ox}$, of the passive layer is comparable to the electrode thicknesses. Table I reports the relevant electric parameters (at $4.2\,$K) for the $Nb/Al_{ox}/Nb$ and $Nb/AlN/NbN$ Josephson tunnel junctions used for this work. For all samples it was $\rm{L}>>\lambda'_j<W$. All measurements were carried out at $4.2\,$K. The experimental set-up has already been described elsewhere\cite{PRB12}.

\begin{table}[tb]
\begin{tabular*}{0.48\textwidth}{@{\extracolsep{\fill}}lcccccccccc}
\hline
\hline
Sample & $J_c$ & $d_i$ & $d_m$ & $d_j$ & $\lambda_J$ & $H_{c}$ & $d'_m$ & $d'_j$ & $\lambda'_J$ & $H'_{c}$ \\
 & $kA/cm^2$ & $nm$ & $nm$ & $nm$ & $\mu m$ &  $A/m$& $nm$ & $nm$ & $\mu m$ &  $A/m$\\
\hline
$Nb$-$Nb$ & $11$  & $400$ & $160$ & $180$ & $3.7$ & $890$ & $380$  & $89$ & $5.3$ & $260$\\
$Nb$-$NbN$  & $3.6$ & $780$ & $270$  & $530$ & $3.6$ & $520$ & $470$  & $210$ & $5.9$ & $190$\\
\hline
\hline
\end{tabular*}
\caption{Relevant electric parameters (at $4.2\,$K) for the $Nb/Al_{ox}/Nb$ and $Nb/AlN/NbN$ Josephson tunnel junctions. The values are approximated to two significant digits and the prime symbol (') denotes the parameters relative to the window junctions. All samples have the same length, $\rm{L}=100\, \mu m$, width, $W=1.2\,\mu m$, and idle region width, $W_i=2.8\,\mu m$.}
\end{table}

\subsection{Magnetic diffraction patterns}

\begin{figure}[tb]
\centering
\subfigure[ ]{\includegraphics[width=7cm]{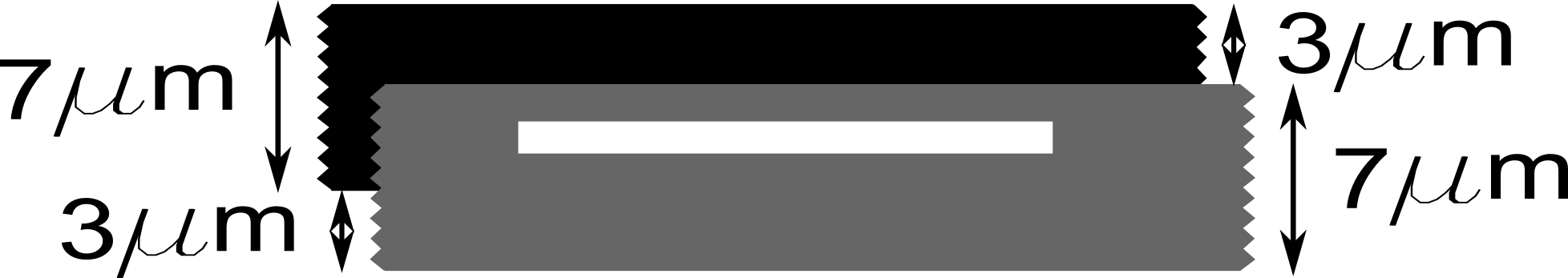}}
\subfigure[ ]{\includegraphics[width=7cm]{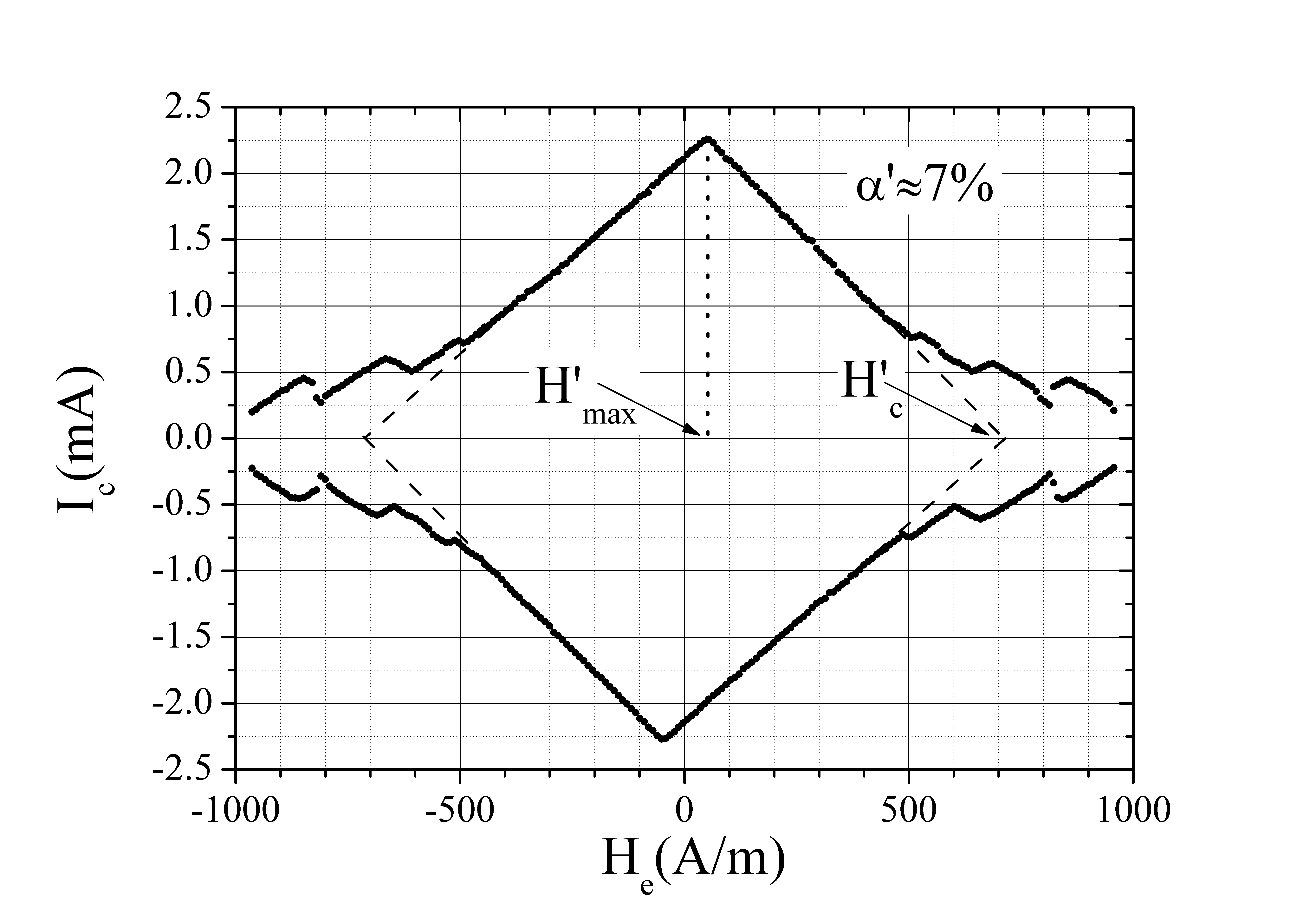}}
\subfigure[ ]{\includegraphics[width=7cm]{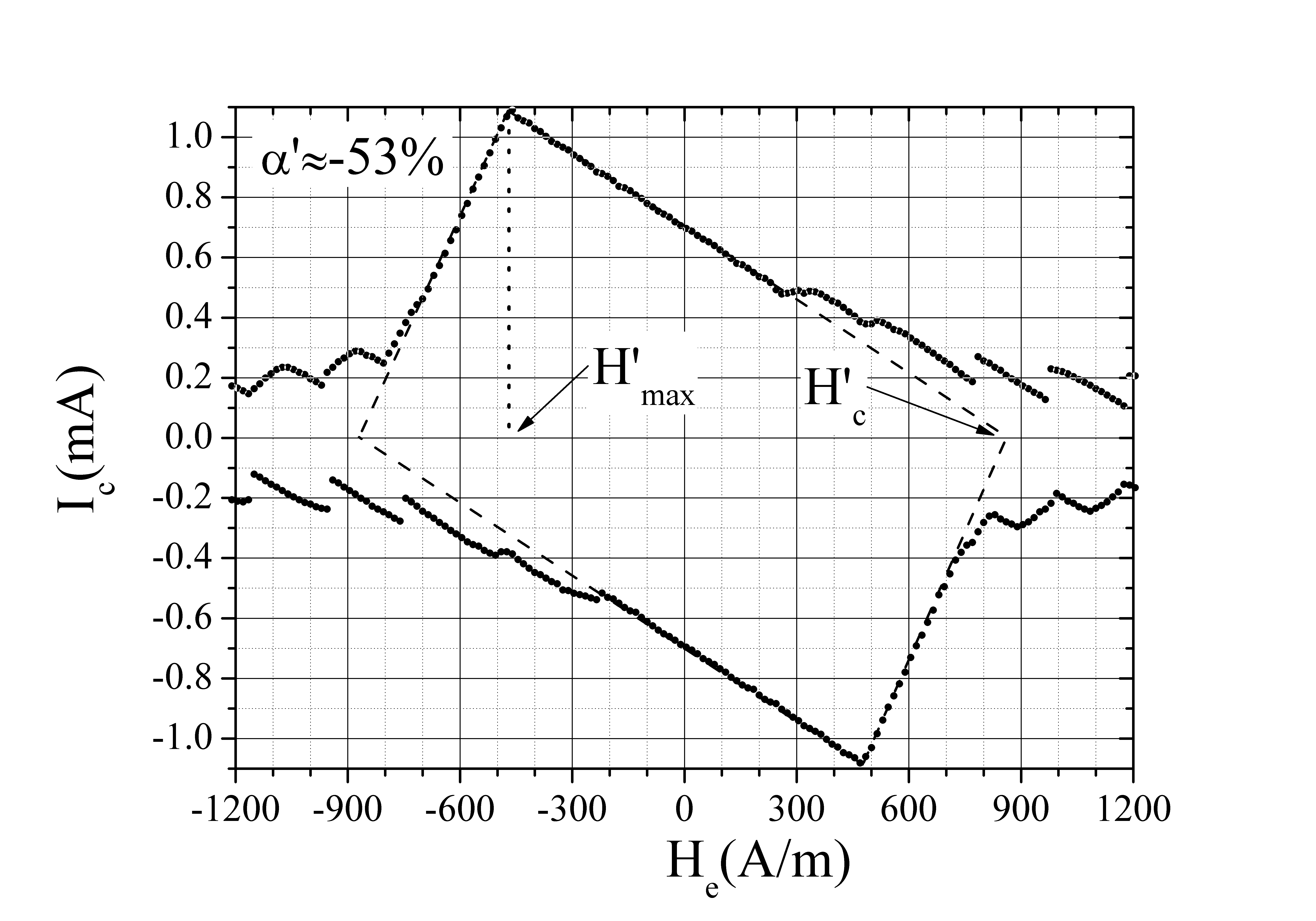}}
\caption{Experimental magnetic diffraction patterns, $I_c(H_{e})$, for two symmetrically biased in-line long Josephson tunnel junctions having equal width electrodes ($W_b=W_t=7 \,\mu m$), but made of different materials: (a) electrode configuration; (b) $Nb$-$Nb$ sample: $I_{c,max}=2.28\, mA$, $H_{max}=50\, A/m$ and $H_{c}=\pm 710\, A/m$ and (c) $Nb$-$NbN$ sample: $I_{c,max}=1.08\, mA$, $H_{max}=-465\, A/m$ and $H_{c}=\pm 870\, A/m$.}
\label{MDP1}
\end{figure}

The upper panel of Fig.~\ref{MDP1} shows the geometry of a in-line junction realized by equal width electrodes, $W_b=W_t=7 \, \mu m$. The electrodes are vertically shifted, but they preserve the symmetry with respect to the junction axis. In Fig.~\ref{MDP1}(b) we report the MDP of such junction when fabricated with the all-$Nb$ technology. The small positive asymmetry, $\alpha'=H'_{max}/H'_c \approx 7\%$, can be fully ascribed to the difference in the electrode thicknesses, $d_w \simeq 2.5\,d_b$, and implies that $\mathcal{L'}_t$ is sligthly smaller than $\mathcal{L'}_b$. From Eqs.(\ref{ratioth}) with $\sigma=1$, we have $\mathcal{L'}_t/\mathcal{L'}_b= 0.90$, in quite good agreement with the value of about $0.88$ obtained from Eq.(\ref{ratioex}). Fig.~\ref{MDP1}(c) is the counterpart of Fig.~\ref{MDP1}(b) for a $Nb$-$NbN$ junction having the same geometry. Now the asymmetry parameter, $\alpha' \approx -0.53$, is negative indicating that $\mathcal{L'}_t > \mathcal{L'}_b$. The factor $\sigma$ in Eqs.(\ref{ratioth}) must be set to $1.9$ to  reproduce the measured ratio $\mathcal{L'}_t/\mathcal{L'}_b=3.3$. 

\begin{figure}[tb]
    \centering
    \subfigure[ ]{\includegraphics[width=7cm]{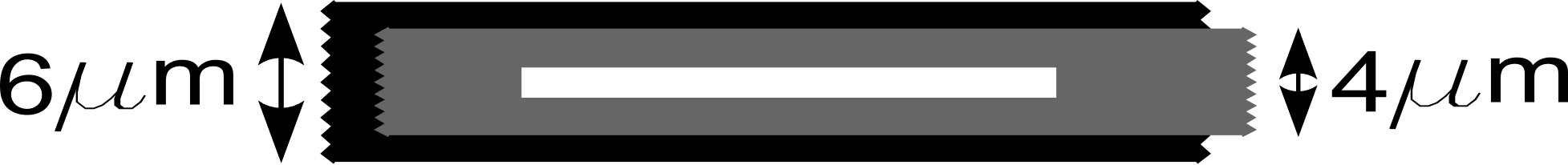}}
    \subfigure[ ]{\includegraphics[width=7cm]{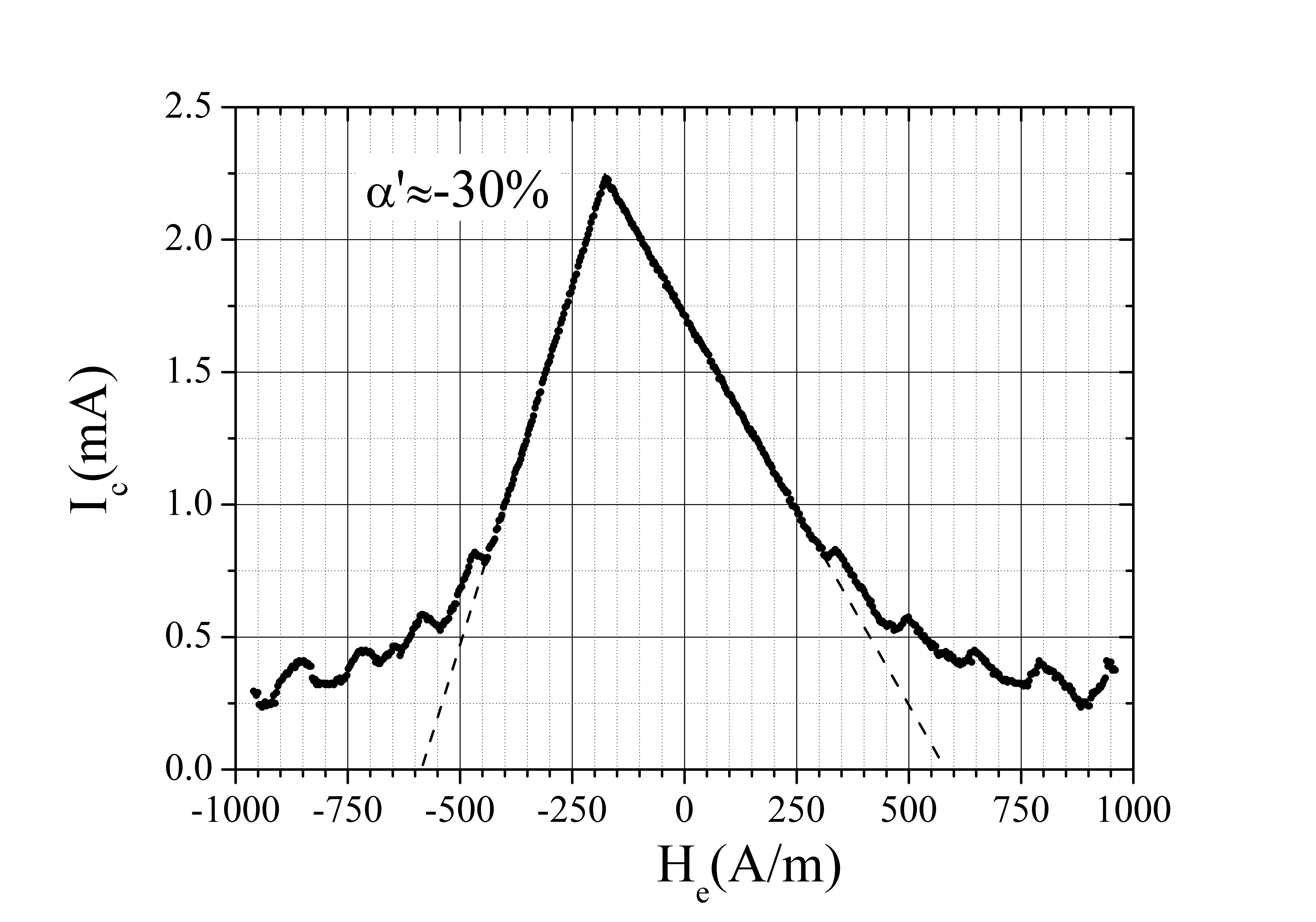}}
    \subfigure[ ]{\includegraphics[width=7cm]{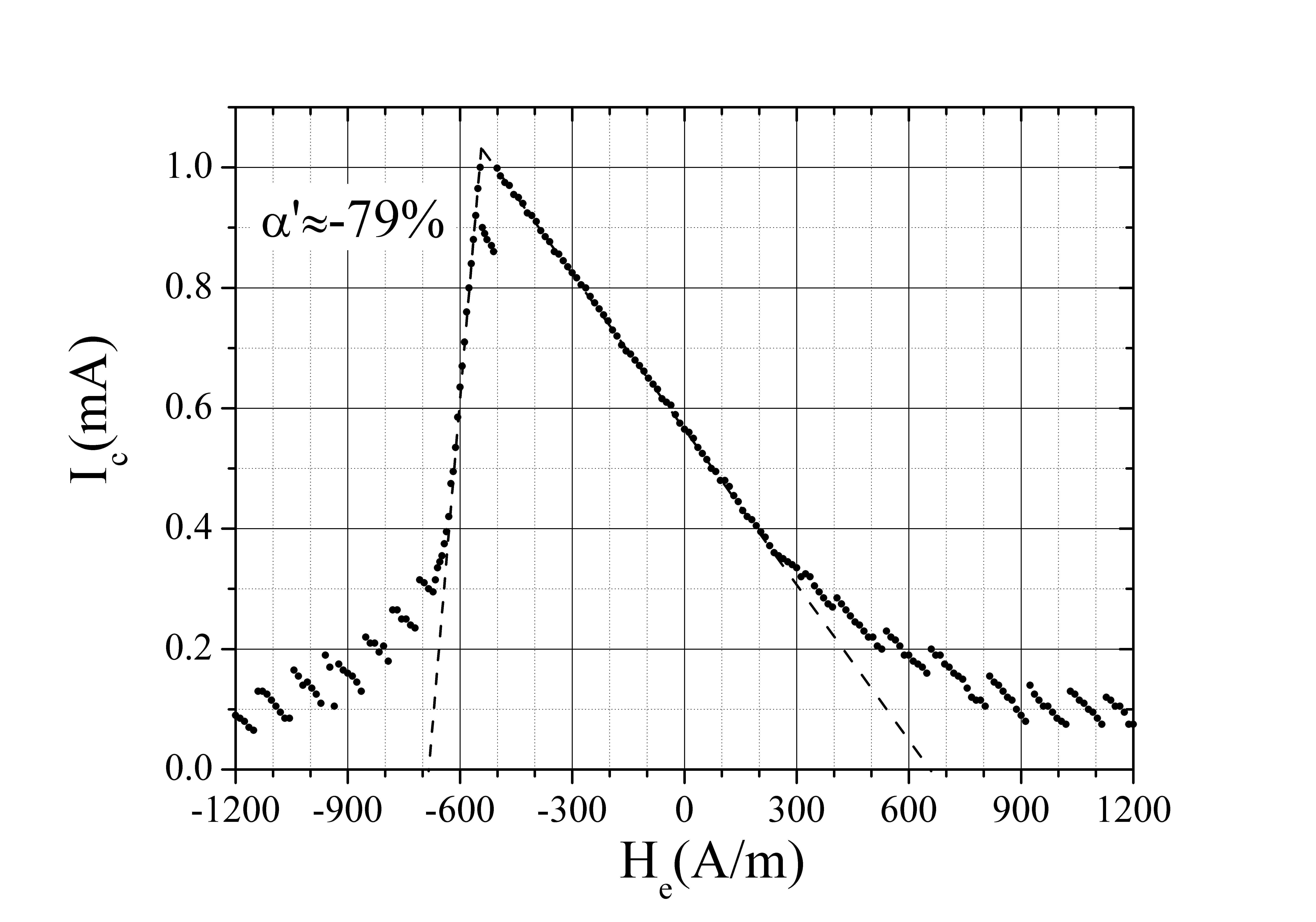}}
    \caption {Experimental magnetic diffraction patterns, $I_c(H_{e})$, for two symmetrically biased in-line long Josephson tunnel junctions having the same geometrical configuration ($W_t=4 \,\mu m$ and $W_b=6 \,\mu m$), but made of different materials: (a) electrode configuration; (b) $Nb$-$Nb$; $I_{c,max}=2.23\, mA$, $H_{max}=-175\, A/m$ and $H_{c}=\pm 585\, A/m$ and (c) $Nb$-$NbN$; $I_{c,max}=1.03\, mA$, $H_{max}=-545\, A/m$ and $H_{c}=\pm 680\, A/m$.}
\label{MDP2}
\end{figure}

\noindent Figs.~\ref{MDP2}(b) and (c) compare the MDPs of two symmetrically biased junctions having the geometrical configuration, depicted in Fig.~\ref{MDP2}(a) ($W_b=1.5W_t=6 \, \mu m$), but made by different materials, respectively, $Nb$-$Nb$ and $Nb$-$NbN$. We now have asymmetries of $\alpha'\approx -30\%$ and $-89\%$, respectively. For the former sample the asymmetry is mainly ascribed to the difference in the electrode widhts, while for the latter it is further enhanced by the large penetration depth of the upper ($NbN$) electrode. To reproduce these values, we had to set $\sigma$ in Eqs.(\ref{ratioth}) to $1.4$ and $3.3$, respectively.  From Eqs.(\ref{ratioex}) we have $\mathcal{L'}_{t,NbN}\approx 4.6 \mathcal{L'}_{t,Nb} \approx  8.5 \mathcal{L'}_b$.

\subsection{Current diffraction patterns}

\begin{figure}[tb]
    \centering
    \subfigure[ ]{\includegraphics[width=7cm]{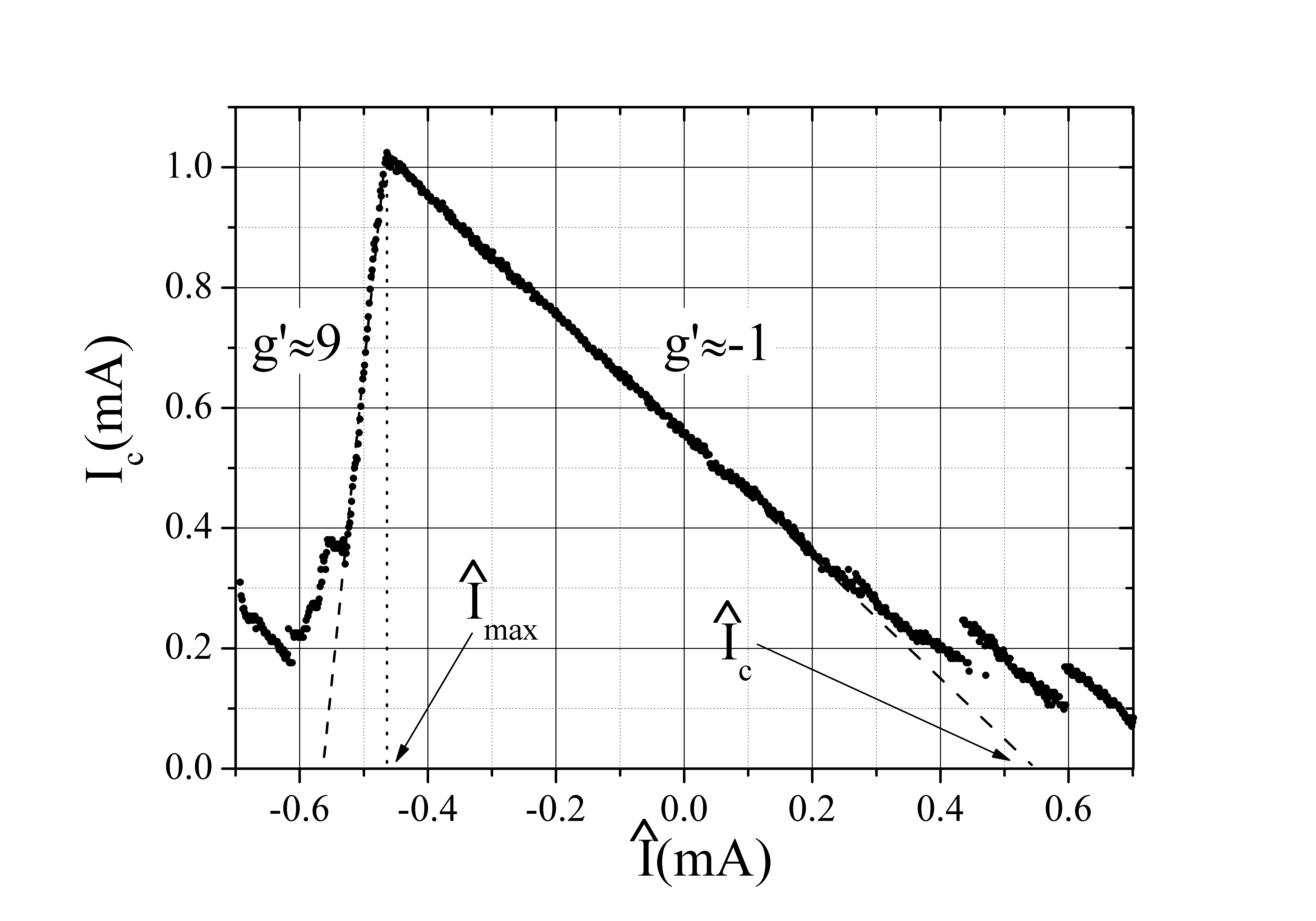}}
    \subfigure[ ]{\includegraphics[width=7cm]{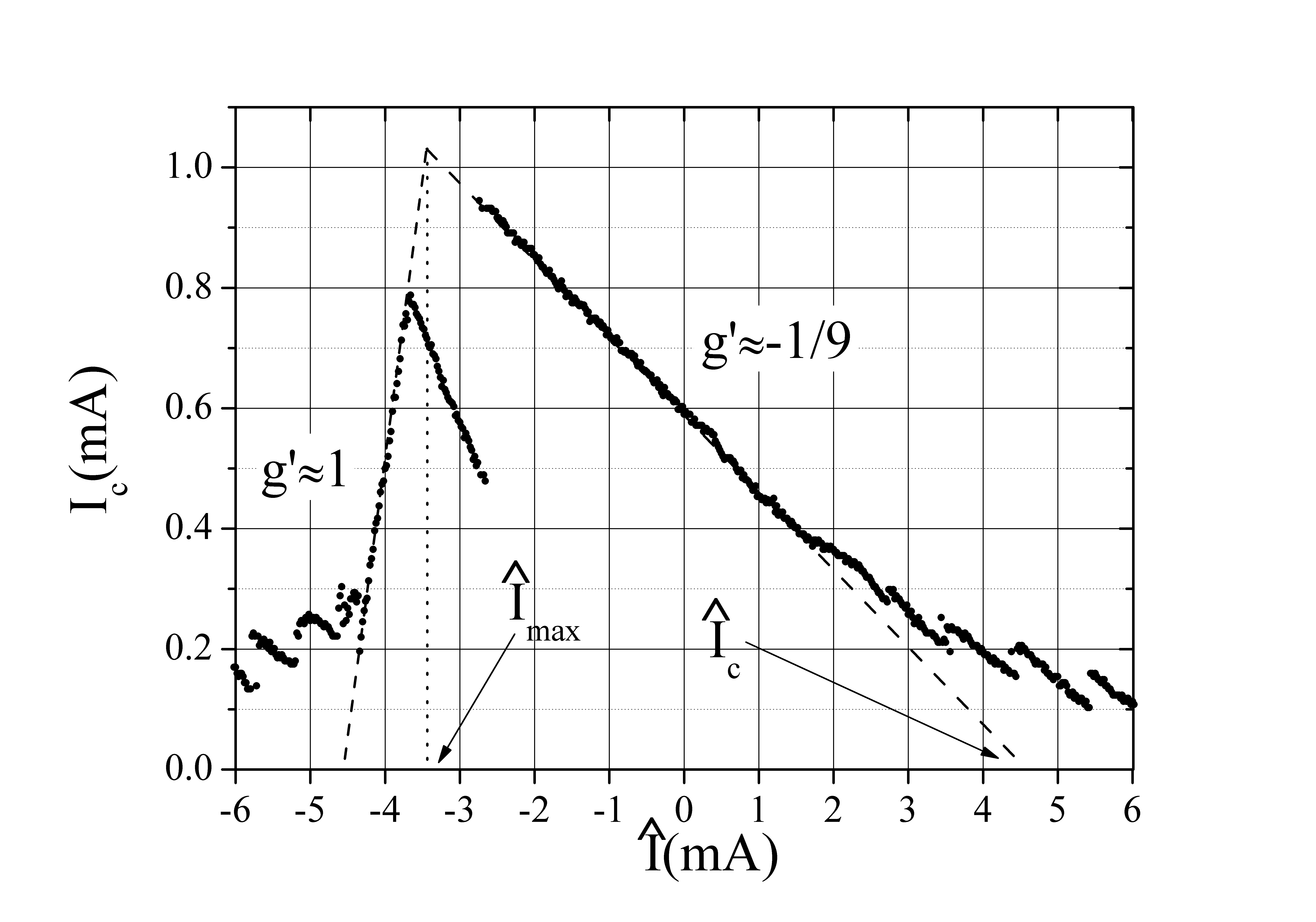}}
    \caption{Experimental current diffraction patterns for the same sample of Fig.~\ref{MDP2}(c). The control current $\hat{I}$ is injected: (a) into the top electrode: $I_{c,max}=1.03\, mA$, $\hat{I}_{max}=-0.46\, mA$ and $\hat{I}_c=\pm 0.56\, mA$, (b) into the base electrode: $I_{c,max}=1.03\, mA$, $\hat{I}_{max}=-3.4\, mA$ and $\hat{I}_c=\pm 4.6\, mA$. $g'$ is the current gain given in Eqs.(\ref{gain'}) or (\ref{gain''}).}
\label{CDP1}
\end{figure}

Fig.~\ref{CDP1}(a) shows the CDP of the same $Nb$-$NbN$ sample of Fig.~\ref{MDP2}(c) when the control current flows in the top (wiring) electrode and in absence of an externally applied field. Comparing it to the MDP in Fig.~\ref{MDP2}(c), we observe a similar qualitative behavior and, as expected according to Eq.(\ref{Icc}), the measured asymmetry parameters are the same within the experimental uncertainty of a few percent. Furthermore the current gains, $g'$, measured by the slopes of the left and right branches of the Meissner lobe are in agreement with the expectation of Eq.(\ref{gain'}).

\noindent If the control current is fed to the bottom, rather than the top electrode, the normalized inductance per unit length of the junction bottom electrode, $\Lambda'_b$, must replace $\Lambda'_t$ in the numerator of the fractions in Eq.(\ref{mdpcl'}), so that

\begin{equation}
\label{gain''}
g'=
\begin{cases} 
			1   & \mbox{for $-\hat{\iota}'_{c} \leq \hat{\iota}' \leq \hat{\iota}'_{max}$}\\
			-\frac{\mathcal{L'}_b}{\mathcal{L'}_t} & \mbox{ for $\hat{\iota}'_{max} \leq \hat{\iota}' \leq \hat{\iota}'_{c}.$}\\
\end{cases}
\end{equation}

\noindent where now $\hat{\iota}'_{c} \equiv 2/\Lambda'_b$  and $\hat{\iota}'_{max}= 2(\Lambda'_t/\Lambda'_b -1)/(\Lambda'_b+\Lambda'_t)$. Fig.~\ref{CDP1}(b) is the CDP of the same sample when the control current is injected into the base electrode. The asymmetry parameter is still unchanged and the pattern slopes agree with Eq.(\ref{gain''}). Other results in agreement with the theory (and not reported here) where obtained for the $Nb$-$Nb$ sample of Fig.~\ref{MDP2}(a). The wide range of linearity of the CDPs is very attractive for the realization of cryogenic current amplifiers with a large dynamic range especially because large slopes can be achieved in the $[-\hat{\iota}_{c},\hat{\iota}_{max}]$ interval. The asymmetry in the electrode inductance can help to significantly improve the gain of a current amplifier over a device with symmetric inductances. The most practical way to achieve this is to inject the signal current in the electrode having the largest inductance per unit length. In practice, the top one is chosen which can be made thin, narrow and of a material with a large penetration depth, such as polycrystalline $NbN$; possibly, on the contrary, the base electrode should be thick, moderately wider and made of a material having a low penetration depth, such as epitaxial $Nb$.

\subsection{Comments}

\noindent As a general qualitative comment to the experimental patterns reported so far, it is worth to mention that the measured critical magnetic fields, $H'_c$, reported in the figure captions, are systematically higher than those in Table I expected from Eq.(\ref{Hc'}) ($2$-$3$ times for the $Nb$-$Nb$ samples and $3$-$5$ times for the $Nb$-$NbN$ ones). One might conclude that the expression for $H'_c$ is wrong or that the parameters used for the calculation are unreasonable. Nevertheless, when comparing the measured and expected values, $\Phi_0/\pi \mathcal{L'}_{t,b} \lambda'_J =\mu_0 H'_c d'_m / \mathcal{L'}_{t,b}$, of the critical control currents, we found a quantitative agreement within the experimental uncertainties. In fact, setting $K_t=1$ in Eq.(\ref{L't}) and $K_b=\sigma=3.3$ in Eq.(\ref{L'b}), we expect the top and bottom critical control currents to be $0.55$ and $4.7\, mA$, respectively; these values have to be compared with those found in the experiments and stated in the caption of Fig.~\ref{CDP1}. We believe that the expression for the critical magnetic field is correct, but, due to demagnetization effects, the externally applied magnetic field, $H_e$, is partially screened by the electrodes\cite{PRB12}. Of course, the current-induced magnetic fields do not suffer from such screening. Therefore, a systematic investigation of magnetic and current diffraction patterns of symmetrically biased in-line LJTJs could be useful to understand the demagnetization effects in superconducting thin-film structures in presence of an in-plane external magnetic field. We observe that the inductance ratio (and so the asymmetry parameter) can also be extracted from the ratio of the top and bottom critical control currents. 

\subsection{Flux diffraction patterns}

\begin{figure}[tb]
    \centering
    \includegraphics[width=6cm]{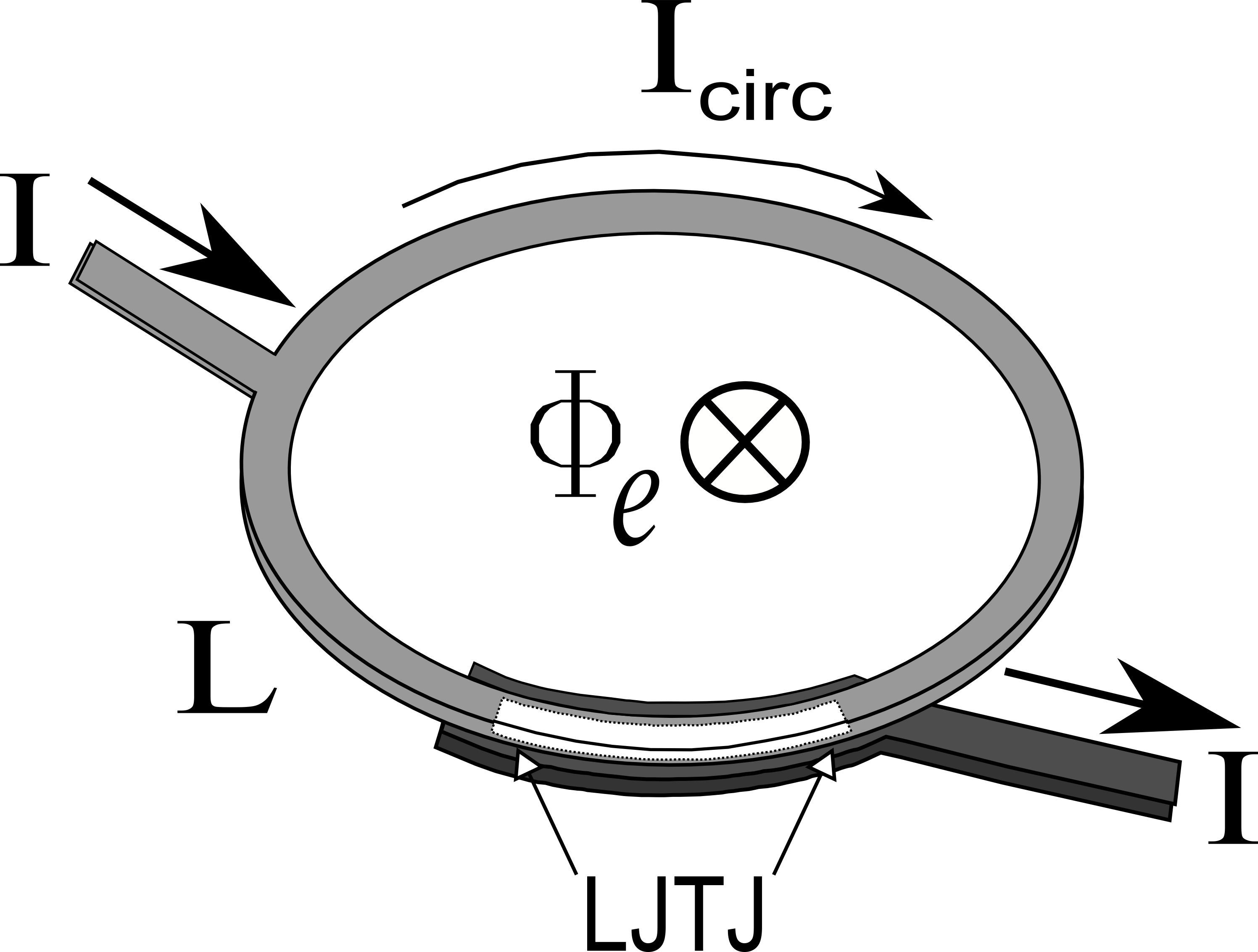}
    \caption{Schematic view of a long Josephson tunnel junction having a ring-shaped  top electrode (in gray); the base electrode is in black and the tunnel area is white.}
\label{DOCELJTJ}
\end{figure}

\begin{figure}[tb]
    \centering
    \includegraphics[width=7cm]{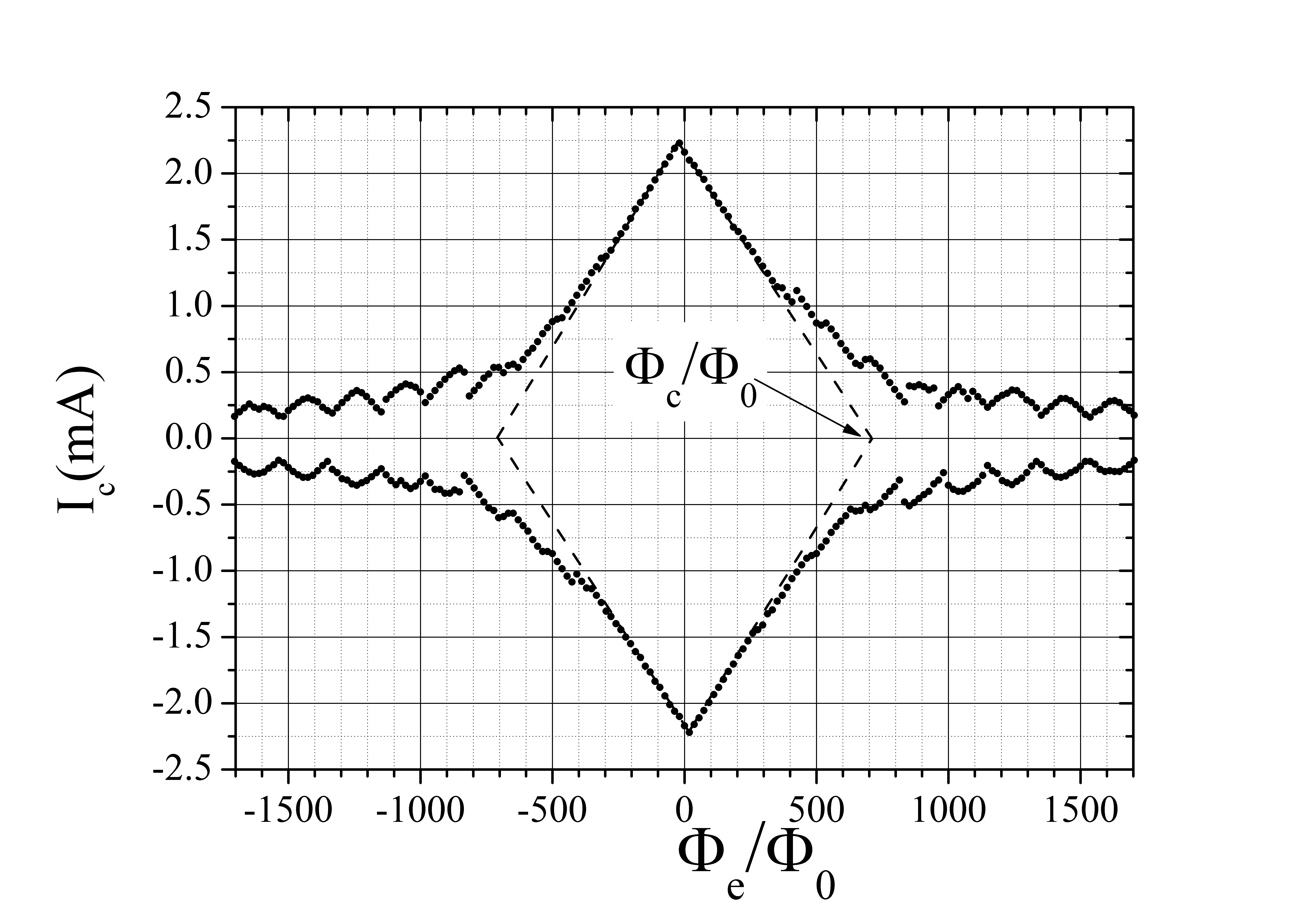}
    \caption{Experimental flux diffraction patterns, $I_c(\Phi_e)$. The shielding currents induced by the external flux circulate in the top electrode: $I_{c,max}=2.25\, mA$ and $\Phi_c=\pm 700 \Phi_0$.}
\label{fdp}
\end{figure}

In a superconducting loop immersed in a magnetic field a current, $\hat{I}$, circulates to expel the field from the loop hole. Denoting with $\Phi_e$ the magnetic flux through the hole and with $L$ the loop geometric (temperature-independent) inductance, then $\hat{I}=\Phi_e/L$. When the loop thickness, $d$, or width, $w$, are comparable with the penetration depth the magnetic flux has to be replaced by the London fluxoid\cite{london} and, at the same time, the kinetic inductance\cite{brandt}, $\mu_0 \lambda^2 /wd$, must be added to the geometric inductance. The circulating current can be detected if, as shown in Fig.~\ref{DOCELJTJ}, a portion of the loop also acts as the electrode of an in-line LJTJ for which $\hat{I}$ is \textit{seen} as a control current capable to modulate its critical current. This mechanism has been recently proposed and demonstrated with the aim to realize magnetic sensors based on LJTJs\cite{SUST12}. Apart from their potential applications, the interest for LJTJs built on a superconducting loop stems from the fact that they were also successfully used to detect trapped flux quanta in a cosmological experiment aimed to study the spontaneous defect production during the fast quenching of a superconducting loop through its normal to superconducting transition temperature\cite{PRBr09}. Fig.~\ref{fdp} shows the dependence of the junction critical current, $I_c$, on the external flux $\Phi_e$, through a $Nb$ ring of inner radius $200\,\mu m$, width $4\,\mu m$ and thickness $470\, nm$. The magnetic flux was applied by means of a calibrated multi-turn coil placed underneath the chip holder whose axis was perpendicular to the loop plane. The pattern asymmetry is very small ($\alpha<1\%$) indicating that the boundary conditions at the junction extremeties are pretty the same; this is made possible by the splitting of the bias currents in the two arms of the loop. The static properties of LJTJs with doubly connected electrodes have been recently investigated\cite{PRB12}. Since the junction critical field only depends on the barrier parameters and not on the electrode configuration, in keeping with our previous findings, it is easy to derive that the critical magnetic flux, $\Phi_c$, is proportional to the loop inductance, $L$, and

\begin{equation}
L=\pi \lambda'_J \mathcal{L'}_t \frac{\Phi_c}{\Phi_0};
\label{L}
\end{equation}

\noindent clearly, it is assumed that the presence of the tunnel junction does not change significantly the inductance of the loop. Eq.(\ref{L}) can be exploited to determine the loop inductance in all those cases where the loop has an irregular geometry or is affected by the presence of other superconducting elements. We have successfully tested Eq.(\ref{L}) for several circular and rectangular loops with different dimensions. In all cases a small portion of the loops constituted the top electrode of a LJTJ whose base electrode was made by a larger $Nb$ film.  As expected, no significant difference was found between the inductance values obtained for $Nb$ and $NbN$ loops. This made us confident with the reliability of the whole experimental procedure and with the values chosen for the bulk magnetic penetrations of these two materials.

\section{Conclusions}

We have addressed the issue of the self-field effects in long Josephson tunnel junctions which were traditionally used to investigate the physics of non-linear phenomena\cite{barone}. In all previous works on LJTJs it was implicitly assumed that the junction electrodes had the same inductance per unit length. Following Weinhact we have generalized the conditions at the junction boundaries for those more realistic cases in which the electrode widths are the same, but their thicknesses and materials are unlike. This case requires the separated knowledge of $\mathcal{L}_b$ and $\mathcal{L}_t$, the inductances per unit length of, respectively, the base and top electrode, related to the magnetic energy stored within a London penetration distance of the film inner surfaces. One interesting feature is that the inductance ratio is directly related to the so far unexplained asymmetry in the magnetic diffraction pattern of symmetrically biased in-line junctions. Later on the modeling was extended to the more common situation in which also the electrode widths can be different and useful expressions for $\mathcal{L}_b$ and $\mathcal{L}_t$ have been proposed, as far as the film widths are not much wider than the tunnel barrier. Our approach also include junctions with a mixed in-line and overlap bias configuration. We like to stress that our analysis on windows-type junctions is restricted to the cases when the film widths are larger than, but comparable with, the junction width. We have reported an extensive experimental study of the static properties of long in-line junctions having different materials and various geometrical configurations. One more interesting feature was that the junction critical current is equally modulated by an external magnetic field or a control current injected into any of the electrodes. The same behavior is reproduced in presence of a magnetic flux linked to a doubly connected electrode. We have shown that our experimental data are compatible with Eqs.(\ref{L't}) and (b) that have to be taken into consideration in the assessment of the self-field effects in Josephson devices.

\section*{Acknowledgments}

\noindent AM and VPK acknowledge the  financial support from the Russian Foundation for Basic Research and the Ministry of Education and Science of the Russian Federation.

\newpage

\end{document}